\documentclass[sigconf]{acmart}
\usepackage{algorithmic}
\usepackage{multirow}
\usepackage[ruled]{algorithm2e}
\usepackage{subfigure}

\copyrightyear{2021} 
\acmYear{2021} 
\setcopyright{acmcopyright}\acmConference[RecSys '21]{Fifteenth ACM Conference on Recommender Systems}{September 27-October 1, 2021}{Amsterdam, Netherlands}
\acmBooktitle{Fifteenth ACM Conference on Recommender Systems (RecSys '21), September 27-October 1, 2021, Amsterdam, Netherlands}
\acmPrice{15.00}
\acmDOI{10.1145/3460231.3474239}
\acmISBN{978-1-4503-8458-2/21/09}

\begin{document}
\title{Learning an Adaptive Meta Model-Generator for Incrementally Updating Recommender Systems}


\author{Danni Peng}
\email{danni001@ntu.edu.sg}
\affiliation{
  \institution{Nanyang Technological University}
  \country{Singapore}
}

\author{Sinno Jialin Pan}
\email{sinnopan@ntu.edu.sg}
\affiliation{
  \institution{Nanyang Technological University}
  \country{Singapore}
}

\author{Jie Zhang}
\email{zhangj@ntu.edu.sg}
\affiliation{
  \institution{Nanyang Technological University}
  \country{Singapore}
}

\author{Anxiang Zeng}
\email{zeng0118@ntu.edu.sg}
\affiliation{
  \institution{Nanyang Technological University}
  \country{Singapore}
}
\begin{abstract}
Recommender Systems (RSs) in real-world applications often deal with billions of user interactions daily. To capture the most recent trends effectively, it is common to update the model incrementally using only the newly arrived data. However, this may impede the model's ability to retain long-term information due to the potential overfitting and forgetting issues. To address this problem, we propose a novel \textbf
{A}daptive \textbf{S}equential \textbf{M}odel \textbf{G}eneration (ASMG) framework, which generates a better serving model from a sequence of historical models via a meta generator. For the design of the meta generator, we propose to employ Gated Recurrent Units (GRUs) to leverage its ability to capture the long-term dependencies. We further introduce some novel strategies to apply together with the GRU meta generator, which not only improve its computational efficiency but also enable more accurate sequential modeling. By instantiating the model-agnostic framework on a general deep learning-based RS model, we demonstrate that our method achieves state-of-the-art performance on three public datasets and one industrial dataset.
\end{abstract}



\maketitle
\section{Introduction}
Recommender Systems (RSs) serve to identify the products or services that could be of  potential interest to users by leveraging personal interaction records and various contextual information. With the prevalence of deep learning in recent years, sophisticated deep models are also adopted in RSs owing to their outstanding performance. As the RS models become increasingly complex and cumbersome, how to update them efficiently and effectively in a data-streaming environment (which is often the case in real-world applications) has emerged as a challenge \cite{wang2020practical,zhang2020retrain}.\

To capture the most recent interest drift of user and change in item popularity, it is important to periodically update the model as new data arrives. Various methods have been proposed to update conventional RS algorithms in the streaming setting \cite{rendle2008online,devooght2015dynamic,vinagre2014fast,papagelis2005incremental,wang2016incremental}. From the perspective of data utilization, the model can be updated using only the newly arrived data (termed \textbf{incremental update}) or using the most recent $w$ periods of data (termed \textbf{batch update}), where $w$ is the size of the sliding window \cite{mi2020ader,wang2020practical,mi2020memory}. In practice, incremental update is often preferred over batch update. Apart from the more efficient training, incremental update can also result in better performance most of the time, as training exclusively with the most recent data can help capture the fast-changing trends and short-term user interests more accurately, which are highly valuable for the near future prediction \cite{chen2013terec,jugovac2018streamingrec,frigo2017online}. \

However, there exists the problem of forgetting for incremental update, which limits the performance of models updated in this manner \cite{diaz2012real,mi2020memory,wang2018streaming,zhao2020stratified}. Since the model is only trained to fit the data from the current period, it is difficult for it to retain and consolidate the long-term memory, as the past knowledge learned is constantly overwritten by the new knowledge representing a distinct distribution. In fact, the forgetting issue is not exclusive to RS incremental update. It generally exists when updating neural networks with data from a different distribution \cite{kirkpatrick2017overcoming}. Continual learning is a field of study that specifically tackles this problem, which has developed effective methods for applications in Computer Vision (CV) and Natural Language Processing (NLP). However, when it comes to RSs incremental update, directly adopting these methods may not lead to desirable outcomes. This is because the two problems have different ultimate goals: continual learning aims to perform well for the current task without sacrificing performance on previous tasks, while incremental update of RSs only cares about the performance on future tasks \cite{zhang2020retrain}. Hence, the main focus of incremental update lies in how to effectively transfer past knowledge especially useful for future predictions.\

To overcome the specific forgetting issue in RSs incremental update, two lines of research have been developed inspired by some of the methods in continual learning \cite{mi2020ader}, namely \textbf{sample-based approach} and \textbf{model-based approach}. Despite the effectiveness of both approaches, there exist some major limitations. For the sample-based approach, although the individual samples are carefully selected to be the `informative' ones, they can hardly represent the big picture of the overall distribution. To tackle this, the model-based approach which considers transfer of knowledge between past and present models was recently proposed. However, a major limitation of the existing model-based methods is that, none of them explores the potential of the long-term sequential patterns exist in model evolution, which can be very useful information for generating a better model for future serving.\

Inspired by this, we propose a novel \textbf{A}daptive \textbf{S}equential \textbf{M}odel \textbf{G}eneration (ASMG) framework for incrementally updating RSs, which encodes a sequence of historical models to generate a better up-to-date serving model via a meta generator. The meta generator is adaptively trained to optimize for the next period data to extract past knowledge that is specially useful for future serving. For the design of the meta generator, we propose to employ Gated Recurrent Units (GRUs) to leverage its ability to capture long-term dependencies. We also introduce some novel strategies to train the GRU meta generator, which not only improve the training efficiency, but also enable better sequential modeling ability. More specifically, we propose to train the GRU meta generator concurrently at multiple steps on the truncated sequence by continuing on a previously learned hidden state. To demonstrate its effectiveness, we instantiate the model-agnostic updating framework on a general deep learning-based Embedding\&MLP model, and conduct extensive experiments compared with state-of-the-art updating methods\footnote{Source code can be found at \url{https://github.com/danni9594/ASMG}.}. Our main contributions are summarized as follows:
\begin{itemize}
\item We propose an ASMG framework which generates a better model by encoding a sequence of historical models.
\item We introduce a GRU-based meta generator design that is capable of capturing the sequential dependencies. We further develop some training strategies to improve its computational efficiency and sequential modeling ability.
\item We demonstrate the effectiveness of our method by conducting extensive experiments on three widely-used public datasets and one industrial production dataset from Lazada.
\end{itemize}

\section{Related Work}
\subsection{Continual Learning}
Methods developed for continual learning generally fall under three categories: experience replay, regularization-based methods, and model fusion. Experience replay prevents forgetting by reusing past samples together with the new data to update the model \cite{robins1995catastrophic}. Generative methods are later on developed to alleviate the burden of storing real data \cite{shin2017continual}. Regularization-based methods retain past knowledge by constraining the parameters update based on some measure of importance \cite{kirkpatrick2017overcoming,zenke2017continual}. Some works also employ distillation techniques to regularize the update direction \cite{rebuffi2017icarl,wen2018few}. Model fusion supports continual learning of tasks by gradually incorporating sub-networks. Both expandable \cite{rusu2016progressive} and fixed-size network methods \cite{mallya2018packnet} are developed. Though these methods have shown promising results in CV and NLP, how to adapt them to incrementally update RS models is non-trivial, as they lack mechanism to explicitly optimize for the future performance. To overcome the specific forgetting problem in RSs incremental update, two lines of research have been developed in recent years, which are closely related to some of the methods in continual learning \cite{mi2020ader}.
\subsection{Sample-based Approach}
This line of works relies on reusing historical samples to preserve past memory, analogous to experience replay in continual learning. Most often, a reservoir is maintained to keep a random sample of history \cite{diaz2012real}. After that, heuristics are designed to select samples from the reservoir for model updating, prioritising recency \cite{chen2013terec,zhao2020stratified} or the extent of being forgetten \cite{guo2019streaming,wang2018streaming,qiu2020gag}. Sample-based approach can be seen as an intermediate between incremental update and training on full historical data, aiming to find a balance between short-term interest and long-term memory. However, an apparent limitation of this approach is that the individual samples, though carefully selected with some measure of informativeness, are insufficient to represent the big picture of overall distribution. In contrast, the historical models learned from the historical data can be seen as a better summarization of the past knowledge. Hence, model-based approach which focuses on transferring knowledge between past and present models was later on developed.
\subsection{Model-based Approach}
This approach aims to overcome forgetting by transferring knowledge between past and present models. Similar to the regularization-based methods in continual learning, \cite{wang2020practical} incorporates a knowledge distillation loss to regularize the model update, using previous incremental model as the teacher model. \cite{xu2020graphsail} and \cite{mi2020ader} also employ the distillation techniques and specifically target at GNN-based and session-based RSs respectively. However, simply constraining the change in networks does not guarantee that the knowledge useful for future prediction is extracted. A recently proposed method termed \textit{Sequential Meta-Learning} (SML) \cite{zhang2020retrain} achieves this by introducing a mechanism to explicitly optimize for the next period data. It devises a transfer module to combine the past and present models, which is adaptively trained in a sequential manner to optimize for future serving. However, the major limitation of both SML and the distillation methods is that they only consider transfer between models of two consecutive incremental periods, ignoring the long-term sequential patterns that may be highly valuable for generating a better model for future serving.\

\section{Methodology}
\begin{figure*}
\vspace{-6mm}
\subfigure[Conventional Incremental Update]{\includegraphics[width=0.41\textwidth]{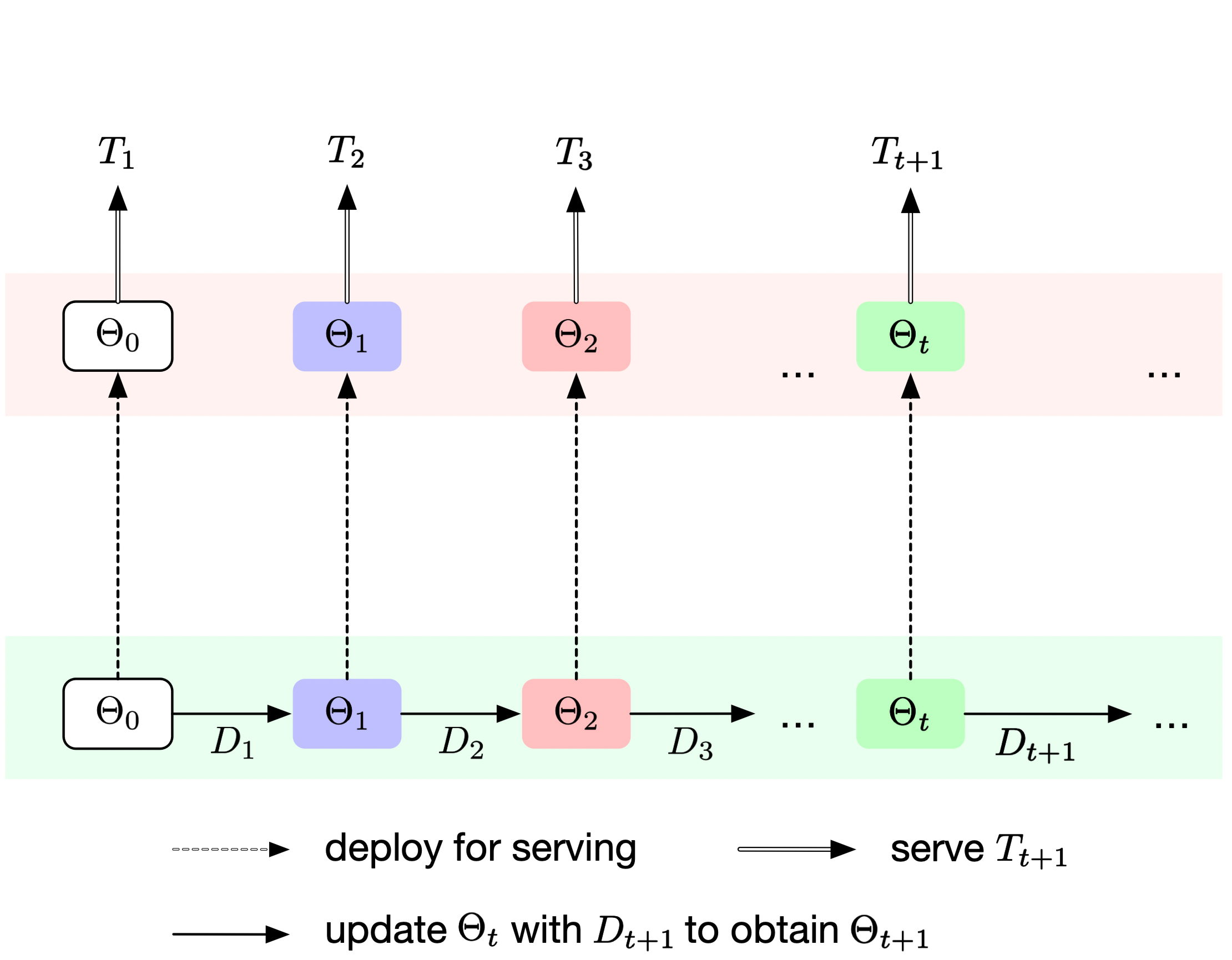}\label{fig:iu}}
\subfigure[ASMG Framework]{\includegraphics[width=0.53\textwidth]{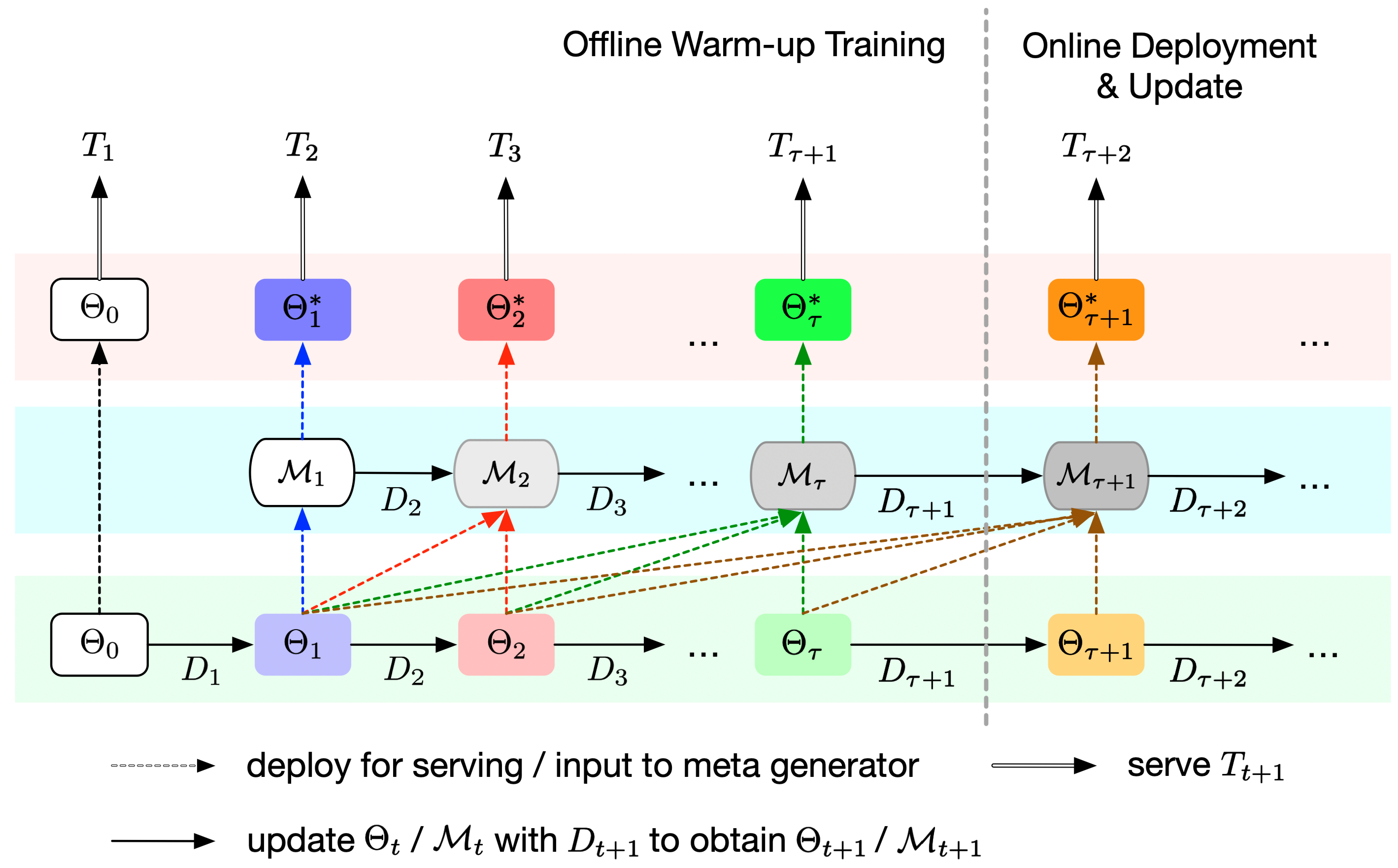}\label{fig:asmg}}
\vspace{-4mm}
\caption{Comparison between conventional incremental update and after applying ASMG framework.}
\vspace{-1mm}
\end{figure*}
\subsection{Problem Description}
Conventional incremental update of RS models involves restoring the last period model as initialization and updating it with the newly arrived data to obtain the most recent model. Formally, let $D_t$ denote the dataset collected from period $T_t$ for $t \in \{1, 2, 3, ...\}$, whereby a period can be a certain length of time (e.g., a day, a week) or until a certain amount of data is collected. The model $\Theta_t$ updated from $\Theta_{t-1}$ is obtained by minimizing loss on $D_t$:
\begin{equation}
\label{eqn:theta:standard}
\Theta_t=\mathop{\arg\min}_{\Theta} \mathcal{L}(\Theta|D_t,\Theta_{t-1}),
\end{equation}
where $\mathcal{L}(\Theta|D_t,\Theta_{t-1})$ is the loss on $D_t$ with $\Theta_{t-1}$ as initialization (for recommendation tasks, the loss can be pointwise log loss or pairwise BPR loss). The initial model $\Theta_0$ can be obtained by random initialization or pre-training with sufficient amount of historical data. Without any post-processing, the updated model $\Theta_t$ is directly deployed to serve for period $T_{t+1}$, from which $D_{t+1}$ can be obtained for training the next period model $\Theta_{t+1}$. The overall procedure of conventional incremental update is illustrated in Figure \ref{fig:iu}.\

Although this updating method is widely adopted due to its efficiency and effectiveness, the model trained in this manner can easily overfit to the current data and forget past patterns learned. 
To address this issue, we propose an \textbf{A}daptive \textbf{S}equential \textbf{M}odel \textbf{G}eneration (ASMG) framework to apply on a sequence of incrementally updated models.\

\subsection{ASMG Framework}
\label{sec:asmg}
For the ASMG framework, instead of direct deployment of the updated model, we introduce a meta generator to take in a sequence of historical models (including the newly updated one), and generate a better model to be deployed for serving the next period. The input sequence of models are obtained from regular incremental update, which means that the model at each period is only trained on the data collected from that period. This allows us to obtain a sequence of models that are highly representative of the respective period. Mining the temporal trends exist in this sequence can have great potential in generating a better model for the next period's serving. Formally, let $\mathcal{M}_t$ denote the meta generator that takes in a sequence of models until $\Theta_t$, the output model $\Theta^*_t$ used for serving $T_{t+1}$ is obtained by:
\begin{equation}
\label{eqn:model_generation}
\Theta^*_t=\mathcal{M}_t(\mathbf{\Theta}_{1:t}),
\end{equation}
where $\mathbf{\Theta}_{1:t}$ is a sequence of models of length $t$ from $\Theta_1$ to $\Theta_t$. Note that $\Theta_0$ is excluded from the input sequence, as it is obtained from random initialization or pre-training instead of regular incremental update. \

The desired properties of an ideal meta generator include 1) good capability of modeling the temporal dependencies exist in the sequence, and 2) stable performance in generating a good model for the next period's serving. The first property is related to the specific network design of the meta generator, which will be discussed in details in the next section. The second property, on the other hand, can be achieved via proper optimization process of the meta generator.\

To generate model that is particularly good at serving the next period, we need to ensure that past knowledge that is specially useful for the next period's serving is extracted. To achieve this, we propose to update the meta generator by adaptively optimizing the output model towards the next period data. More specifically, let $\omega_t$ denote the parameters of $\mathcal{M}_t$, i.e., $\mathcal{M}_t=\mathcal{M}_{\omega_t}$, the parameters $\omega_{t+1}$ are updated from $\omega_t$ by optimizing $\Theta^*_t$ towards $D_{t+1}$:
\begin{equation}
\label{eqn:meta_update}
\begin{aligned}
\omega_{t+1}
& =\mathop{\arg\min}_{\omega} \mathcal{L}(\Theta^*_t|D_{t+1},\omega_t) \\
& =\mathop{\arg\min}_{\omega} \mathcal{L}(\mathcal{M}_\omega(\mathbf{\Theta}_{1:t})|D_{t+1},\omega_t),
\end{aligned}
\end{equation}
where $\mathcal{L}(\Theta^*_t|D_{t+1},\omega_t)$ is the loss on $D_{t+1}$ with $\omega_t$ as initialization. The loss function here is the same as that in \eqref{eqn:theta:standard}.\

To ensure that the meta generator is sufficiently trained before it can be deployed for online model generation, we perform $\tau$ periods of meta generator warm-up training prior to the online deployment. During the online deployment phase, the meta generator will still need to be adaptively updated in order to keep up with the most recent sequential trends. The overall procedure of the ASMG framework is depicted in Figure \ref{fig:asmg} and described in Algorithm \ref{alg:asmg}.

\setlength{\textfloatsep}{3mm}
\begin{algorithm}[t]
\small
\SetKwInput{KwInput}{Input}                
\SetKwInput{KwOutput}{Output}              
\caption{ASMG Framework}
\label{alg:asmg}
\KwInput{Sequence of historical datasets $\{D_t\}_{t=2}^{\tau+1}$ and incrementally updated models $\{\Theta_t\}_{t=1}^{\tau+1}$}
\KwOutput{Trained meta generator $\mathcal{M}_t$ for $t\geq2$, and incrementally updated model $\Theta_t$ for $t\geq\tau+2$}
\textbf{Offline Warm-up Training:}\\
Randomly initialize $\omega_1$\\
\For {$t=1$ to $\tau$}
{$\omega_{t+1} \leftarrow \arg\min_\omega \mathcal{L}(\mathcal{M}_\omega(\mathbf{\Theta}_{1:t})|D_{t+1},\omega_t)$ \ \ // update the meta generator} 
\textbf{Online Deployment \& Update:}\\
\While {$t\geq\tau+1$}
{Generate output model $\Theta^*_t=\mathcal{M}_t(\mathbf{\Theta}_{1:t})$, deploy $\Theta^*_t$ to serve for $T_{t+1}$ and obtain $D_{t+1}$\\
$\Theta_{t+1} \leftarrow \arg\min_\Theta \mathcal{L}(\Theta|D_{t+1},\Theta_t)$ \ \ // update the base model\\
$\omega_{t+1} \leftarrow \arg\min_\omega \mathcal{L}(\mathcal{M}_\omega(\mathbf{\Theta}_{1:t})|D_{t+1},\omega_t)$ \ \ // update the meta generator}
\end{algorithm}

\subsection{GRU Meta Generator}
There are many possible designs for the meta generator. In this work, we propose to employ Gated Recurrent Unit (GRUs) \cite{chung2014empirical} to leverage its ability to capture the sequential patterns. For simplicity, we apply the GRU meta generator coordinate-wise on the base model $\Theta$. That is to say, for each individual $\theta \in \Theta$, we generate the output parameter $\theta^*$ by applying the GRU meta generator on a sequence of its historical values, independently of parameters at all the other coordinates. The final serving model is therefore $\Theta^*=\{{\theta^*}'s\}$.
More specifically, at each step $i \in \{1, ..., t\}$, the hidden state $h_i \in \mathbb{R}^d$ is computed from the last step hidden state $h_{i-1} \in \mathbb{R}^d$ and $\theta \in \mathbb{R}$ of the current step input model $\Theta_i \in \mathbb{R}^n$:
\begin{equation}
\begin{aligned}
& r_i = \sigma(W_r\cdot[h_{i-1},\theta]), \\
& z_i = \sigma(W_z\cdot[h_{i-1},\theta]),  \\
& \tilde{h}_i = \tanh(W_{\tilde{h}}\cdot[r_i  \odot  h_{i-1},\theta]), \\
& h_i= (1-z_i) \odot  h_{i-1}+z_i \odot \tilde{h}_i,\\
\end{aligned}
\label{eqn:gru_cell_1}
\end{equation}
where $r_i,z_i \in \mathbb{R}^d$ are gates that control how much of past and present information to be retained, $W_r,W_z,W_{\tilde{h}} \in \mathbb{R}^{d \times (d+1)}$ are learnable weights, $\sigma(\cdot)$ is the sigmoid function, $[\cdot,\cdot]$ denotes concatenation, and $\odot$ denotes element-wise multiplication. The initial hidden state $h_0$ is zero-initialized.\

The output parameter $\theta^* \in \mathbb{R}$ of the final serving model $\Theta_i^* \in \mathbb{R}^n$ at step $i$ is obtained from the respective hidden state $h_i$ via a linear transformation:
\begin{equation}\label{eqn:gru_cell_2}
\theta^*=W \cdot h_i+b,
\end{equation}
where $W \in \mathbb{R}^{d}$ and $b \in \mathbb{R}$ are also learnable parameters. All the learnable parameters are shared across $t$ steps.

If we assign one unique GRU meta generator (i.e., unshared set of learnable parameters) at each coordinate in the base model, the space complexity of the GRU meta generator will be $\mathcal{O}(nd^2)$ where $d$ is the GRU hidden size and $n$ is the base model parameter size. Note that the size of the GRU meta generator is unrelated to the sequence length.

\subsection{Training Strategies}
In this section, we introduce two training strategies specifically tailored to the GRU meta generator to further improve its training efficiency and sequential modeling ability.

\subsubsection{Training GRU Meta Generator on Truncated Sequence} 
One main drawback of GRUs is that the computation time increases linearly with sequence length. Hence, applying GRU meta generator on the full sequence of historical models is not practical, as the sequence length will continuously increase as time goes by, i.e., the computational complexity is $\mathcal{O}(tnd^2)$, linear in the current time step $t$. An alternative is simply to truncate the sequence and start training from an intermediate step with zero-initialized input hidden state. However, using a zero-initialized input hidden state inevitably discards earlier models and prevents the retention of longer-term information. \

To tackle this problem, we propose to train the GRU meta generator on truncated sequence by continuing on a previously learned hidden state. By doing so, we not only improve training efficiency with shorter sequence, but also preserve long-term memory by reusing a hidden state learned from the past. For consistency, we fix the sequence length at $k$ for some $1 \leq k\leq t$ for all training periods. Specifically, when training $\mathcal{M}_{t+1}$, we take in a sequence of $k$ most recent models $\mathbf{\Theta}_{t-(k-1):t}$ as inputs, and a learned hidden state $h_{t-k}$ obtained from the training of the previous meta generator $\mathcal{M}_t$ as the initial hidden state for the current training. Consistently applying this strategy at every training period endows a nice effect that consolidates all the previous models $\mathbf{\Theta}_{1:t-k}$ into hidden state $h_{t-k}$. Formally, this strategy modifies the way of generating output model in \eqref{eqn:model_generation} to $\Theta^*_t=\mathcal{M}_{t}(\mathbf{\Theta}_{t-(k-1):t},h_{t-k})$. The overall optimization is as follows:
\begin{equation}
\begin{aligned}
\omega_{t+1}
& =\mathop{\arg\min}_\omega \mathcal{L}(\Theta^*_t|D_{t+1},\omega_t) \\
& =\mathop{\arg\min}_{\omega} \mathcal{L}(\mathcal{M}_\omega(\mathbf{\Theta}_{t-(k-1):t},h_{t-k})|D_{t+1},\omega_t).
\label{eqn:strategy_1}
\end{aligned}
\end{equation}
Here, $\mathcal{M}_\omega$ is the GRU meta generator that maps many to one. This strategy is illustrated in Figure \ref{fig:asmg_gru} from (a) to (b).\

\subsubsection{Training GRU Meta Generator at Multiple Steps Concurrently}
\begin{figure*}[t]
\vspace{-1mm}
\centering
\includegraphics[width=1\textwidth]{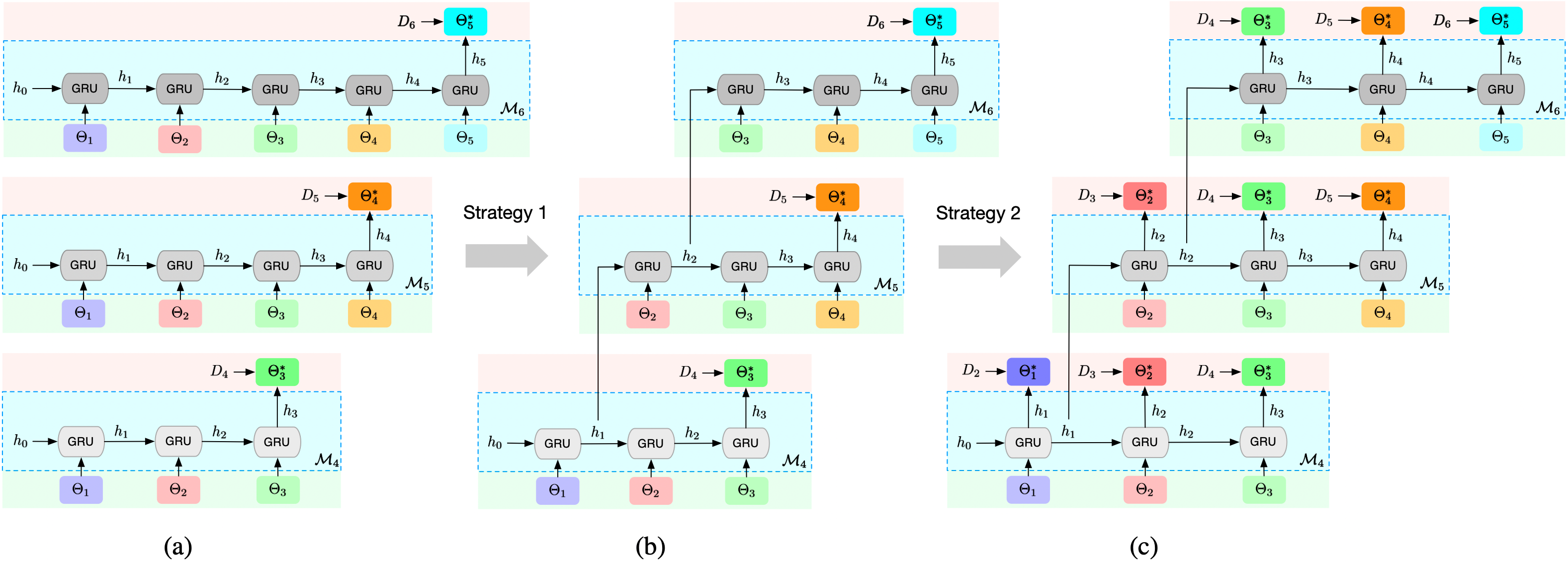}
\vspace{-6mm}
\caption{Training of GRU meta generator under the ASMG framework for three consecutive periods (i.e., training of $\mathcal{M}_4$, $\mathcal{M}_5$ and $\mathcal{M}_6$). Transition from (a) to (b) illustrates the first strategy, which trains the GRU meta generator on truncated sequence (here we use sequence length $k=3$) by continuing on a previously learned hidden state. Transition from (b) to (c) illustrates the second strategy, which performs concurrent training by optimizing data at multiple steps, with greater weights assigned to the more recent data. \textbf{ASMG-GRUsingle} and \textbf{ASMG-GRUmulti} are depicted by (b) and (c) respectively.}
\label{fig:asmg_gru}
\vspace{-1mm}
\end{figure*}

To enable more accurate sequential modeling, we further propose a second strategy to perform concurrent training at multiple steps. The intuition behind is to enforce that the hidden states generated at different steps are on the right track and able to produce model that serves well for the respective next period. More specifically, when training $\mathcal{M}_{t+1}$, instead of optimizing the last model $\Theta^*_t$ towards $D_{t+1}$ only, we optimize all $\{\Theta^*_i\}^t_{i=t-(k-1)}$ towards all $\{D_{i+1}\}^t_{i=t-(k-1)}$ concurrently. To take into account that the later data is less seen before and hence more informative, we assign greater weight $\lambda$ to loss of the more recent data. Further incorporating this strategy results in the following optimization:
\begin{equation}
\begin{aligned}
\omega_{t+1}
& =\mathop{\arg\min}_\omega \!\!\! \sum_{i=t-(k-1)}^t \!\!\! \lambda_i \mathcal{L}(\Theta^*_i|D_{i+1},\omega_t) \\
& =\mathop{\arg\min}_{\omega} \!\!\! \sum_{i=t-(k-1)}^t \!\!\! \lambda_i \mathcal{L}(\mathcal{M}_{\omega}(\mathbf{\Theta}_{t-(k-1):i}, h_{t-k})|D_{i+1},\omega_t),
\label{eqn:strategy_2}
\end{aligned}
\end{equation}
where $\lambda_i$ is the weight for loss $\mathcal{L}_i=\mathcal{L}(\Theta^*_i|D_{i+1},\omega_t)$, computed from a linear decay function, i.e., $\lambda_{t-(k-j)}=\frac{j}{\sum_{j'=1}^kj'}$ for $j \in \{1, 2, ..., k\}$. This strategy is illustrated in Figure \ref{fig:asmg_gru} from (b) to (c).\

Note that for this strategy, we slightly violate the proposed ASMG framework by training the meta generator on data from multiple periods. However, we later on show that this strategy is able to further boost the performance and outperform all the existing model updating methods, including those that also involve multiple periods of data. Furthermore, this strategy should be distinguished from batch update, as the the data from earlier periods are used to train the meta generator instead of the base model. The base model $\Theta_i$, as the inputs to meta generator, are still trained by regular incremental update.\

We term our proposed method that employs GRU meta generator under the ASMG framework as \textbf{ASMG-GRU}. We further differentiate the two strategies by naming the first one that trains on a single period of data as \textbf{ASMG-GRUsingle}, and the second one that trains on multiple periods of data as \textbf{ASMG-GRUmulti}.

\subsection{Instantiation on Embedding\&MLP Base Model}
The proposed ASMG-GRU is model-agnostic. To test its effectiveness, we instantiate it on a general deep learning-based Embedding\&MLP model, a network architecture that most of the Click-Through Rate (CTR) prediction models developed in recent years are based on \cite{shan2016deep,cheng2016wide,guo2017deepfm,zhou2018deep}. The model mainly comprises embedding layers that transform high-dimensional sparse features into low-dimensional dense vectors, and Multi-Layer Perceptron (MLP) layers that learn the interaction of the concatenated feature embeddings. \

More specifically, given user $i$ with $P$ categorical features, the concatenated feature embedding $u_i$ is obtained by:
\begin{equation}
u_i=[E_{u1}z_{i1},\cdots,E_{uP}z_{iP}],
\end{equation}
where $E_{up} \in \mathbb{R}^{d_e \times d_{up}}$ with $d_e \ll d_{up}$ is the embedding matrix for categorical feature $p \in \{1,\cdots,P\}$ of user, and $z_{ip} \in \mathbb{R}^{d_{up}}$ is the one-hot or multi-hot vector for categorical feature $p$ of user $i$.\

Similarly, for item $j$ with $Q$ categorical features, the concatenated feature embedding $v_j$ is obtained by:
\begin{equation}
v_j=[E_{v1}z_{j1},\cdots,E_{vQ}z_{jQ}],
\end{equation}
where $E_{vq} \in \mathbb{R}^{d_e \times d_{vq}}$ with $d_e \ll d_{vq}$ is the embedding matrix for categorical feature $q \in \{1,\cdots,Q\}$ of item, and $z_{jq} \in \mathbb{R}^{d_{vq}}$ is the one-hot or multi-hot vector for categorical feature $q$ of item $j$.\

The predicted preference of user $i$ on item $j$ is then obtained by passing the concatenated feature embeddings through a two-layer MLP, followed by a sigmoid function to normalize the score:
\begin{equation}
\hat{y}=\sigma(\text{MLP}([u_i,v_j])).
\end{equation}
The parameters of the Embedding\&MLP base model $\Theta$ therefore include both the embedding matrices $\{\mathbf{E}_u, \mathbf{E}_v\}$, and the weights and biases of $\text{MLP}(\cdot)$. For CTR prediction, we adopt log loss to update the base model $\Theta$ as described in \eqref{eqn:theta:standard}:
\begin{equation}
\mathcal{L}(\Theta|D_t)=-\frac{1}{|D_t|}\Big(\!\!\sum_{(i,j) \in D^+_t} \!\!\!\! \log(\hat{y})+ \!\!\sum_{(i,j) \in D^-_t} \!\!\!\! \log(1-\hat{y})\Big),
\end{equation}
where $D^-_t$ is obtained by randomly sampling 1 unobserved interaction for each observed interaction in $D^+_t$.

To apply the proposed ASMG-GRU, we assign one unique meta generator to each individual weight and bias parameter in the MLP layers. However, for the embedding layers of sparse ID features (i.e., user ID and item ID), the parameter size is large while the training samples for each ID are limited. Assigning one unique meta generator for each parameter in the embedding layers significantly increases the complexity and can easily lead to overfitting. Hence, we assign one unique meta generator at each dimension of the embedding vector and share it across all IDs. Take for instance, for the $p$-th user embedding matrix $E_{up} \in \mathbb{R}^{d_e \times d_{up}}$, we assign $d_e$ distinct meta generators along the first dimension, and share them across $d_{up}$ IDs along the second dimension. The space complexity of the GRU meta generators for $E_{up}$ will therefore be significantly reduced from $\mathcal{O}(d_ed_{up}d^2)$ to $\mathcal{O}(d_ed^2)$, as $d_{up}$ is often large.\

\section{Experiments}
\subsection{Experimental Settings}
\subsubsection{Datasets}
Experiments are conducted on three widely-used public datasets and one private industrial production dataset collected from Lazada mobile app.

\begin{itemize}
\item {\bf Tmall}\footnote{https://tianchi.aliyun.com/dataset/dataDetail?dataId=42} contains user-item interaction records from Tmall.com. We extract 31-day data from October 2014, splitting into 31 periods by treating each day as a period. We select the top 50,000 users with the most interactions and filter out items with less than 20 interactions. In the end, the pre-processed dataset contains 3,047,101 observed interactions with 49,986 users and 43,571 items from 634 categories.

\item {\bf Sobazaar}\footnote{https://github.com/hainguyen-telenor/Learning-to-rank-from-implicit-feedback} is also an e-commerce dataset that contains user-item interaction records from Sobazaar mobile app from September 2014 to December 2014. We split the 4-month data into 31 periods such that each period has equal amount of data. This results in around 4 days per period. In the end, the pre-processed dataset contains 842,660 observed interactions with 1,7126 users and 2,4785 items.

\item {\bf MovieLens}\footnote{https://grouplens.org/datasets/movielens/} contains explicit user ratings for movies on a scale of 0 to 5. We use the largest 25M dataset and extract data from 2014 to 2018. We split the 5-year data into 31 periods based on equal amount of data, which gives around 2 months per period. For movie recommendations, it makes sense to have each period to span longer time (as compared to the e-commerce domain), as the user's taste in movies is more of a long-term interest and less time-sensitive. To make it suitable for our binary classification base model, we follow \cite{wang2018streaming} and label ratings above 3 as positive, and the rest as negative. In the end, the pre-processed dataset contains 6,840,091 samples (including both positive and negative samples) with 43,181 users and 51,142 movies from 20 genres.

\item {\bf Lazada} is an industrial production dataset collected from Lazada mobile app, the largest e-commerce platform in Southeast Asia. Samples are constructed from the impression logs of users, with clicked records labeled as positive and unclicked records labeled as negative. A 31-day data from December 2020 is extracted and split into 31 periods by treating each day as a period. We randomly select 10,000 users for our experiments. In the end, the pre-processed dataset contains 6,659,828 samples (including both positive and negative samples) with 10,000 users and 43,878 items from 3,860 categories.
\end{itemize}

For all datasets, we collect the most recent 30 positive feedbacks of each user to form the user sequence feature. 



\subsubsection{Baselines}
We compare the following model updating methods applying on the same Embedding\&MLP model.
\begin{itemize}
\item {\bf Incremental Update (IU)}: This method updates the model incrementally using only the new data $D_t$.

\item {\bf Batch Update (BU-w)}: This method updates the model using the most recent $w$ periods of data $\{D_{t-(w-1)}, \cdots, D_t\}$. We experiment the window size $w$ for 3, 5, and 7 periods. Note that IU is equivalent to BU-1.

\item {\bf SPMF} \cite{wang2018streaming}: This method is a sample-based approach. It maintains a reservoir of historical samples, from which the samples with lower predicted ranks are selected to mix with the new data $D_t$ for current model updating. We tune the reservoir size as a fraction of the total dataset size from 0.1 to 0.5.

\item {\bf IncCTR} \cite{wang2020practical}: This method is a model-based approach that incorporates a knowledge distillation loss when training the current model with $D_t$. We implement the version that uses the previous incremental model as teacher model. We tune the weight assigned to the knowledge distillation loss from 0.1 to 0.5.

\item {\bf SML} \cite{zhang2020retrain}: This method is another model-based approach that has recently achieved the state of the art. It employs a CNN-based transfer module to leverage the previous model while training the current model on $D_t$. The transfer of knowledge is only between models of two consecutive periods. SML is originally proposed to instantiate on Matrix Factorization (MF) base model. For a fairer comparison, we also implement this version and term it \textbf{SML-MF}. We tune the number of CNN filters in $\{2, 5, 8, 10\}$, and the MLP hidden size in $\{10, 20, 40, 80\}$.

\item {\bf ASMG-Linear}: To compare with the proposed GRU meta generator design under ASMG framework, we implement another fixed-length meta generator that linearly combines the sequence of models, i.e., $\Theta^*_t=\sum^t_{i=t-(k-1)}\alpha_i\Theta_i$, where $\alpha_i$ is the weight assigned to model at period $i$. We tune the sequence length $k$ from 1 to 6.

\item {\bf ASMG-GRU}: This is our proposed method to employ the GRU meta generator under the ASMG framework. Incorporating the proposed training strategies, we compare \textbf{ASMG-GRUsingle} and \textbf{ASMG-GRUmulti} against the above baselines. We tune the sequence length $k$ from 1 to 6, and the GRU hidden size in $\{4, 8, 12, 16\}$.
\end{itemize}

For all the compared methods, we use Adam optimizer to update  the trainable parameters (i.e., the base model, the meta generator under ASMG, and the transfer module under SML) and tune the learning rate in \{1e-2, 1e-3, 1e-4\}. We also tune the number of epochs for each update period in $\{1, 5, 10, 20\}$. The batch size is set to be 256 for the Sobazaar dataset and 1024 for other datasets.

\subsubsection{Evaluation Protocols}
All datasets are split into 31 periods. The first 10 periods are used to pre-trained an initial model $\Theta_0$. Model updating will be conducted for the remaining 20 periods. As mentioned earlier in section \ref{sec:asmg}, we need to reserve some periods for warm-up training of the meta generator. For input sequence of length $k$, the training of the meta generator can only start from period $10+k$, as we will need to first obtain a sequence of incrementally updated models of length $k$ from period $11$ to $10+k$. Hence, to allow for at least 5 periods of effective warm-up training for a maximum sequence length of $k = 6$ in our experiments, we split the 20 periods into train/validation/test sets by $10/3/7$. Evaluation of the model updated at each test period is based on its prediction performance for the respective next period. The metrics used are \textbf{AUC} (Area Under ROC) and \textbf{LogLoss} (binary cross-entropy), which are good measurements of CTR prediction performance. We conduct 5 runs of model updating experiment for each compared method.

\subsection{Performance Comparison}
Table \ref{tbl:overall_perf} presents the overall performance for the compared methods, from which we have the following observations:
\begin{itemize}
\item BU methods with $w \geq 2$ are not comparable to IU in terms of serving for the next period, and the degradation increases for larger window size. This observation is consistent with \cite{frigo2017online,wang2020practical}, which have identified that in non-stationary environment, batch update is inferior as compared to incremental update. This may be due to that the newly collected data can better represent the latest trends, which are highly related to the upcoming period. Expanding the window size affects the recency of the training data, and hence will be less effective in capturing the latest trends and fast-changing short-term user interests for the near future prediction.
\item SPMF and IncCTR only improve over IU marginally. The former is a sample-based approach. The latter is a model-based approach but only utilizes the historical model indirectly via knowledge distillation. The fact that SML performs the best among all the baseline methods demonstrates that directly incorporating the previous model to derive the transfer patterns can be very effective for generating a better current model. Moreover, the mechanism of SML to train the transfer module by optimizing towards the next period data helps extract knowledge that is particularly useful for future prediction.
\item SML-MF is not comparable to other methods due to a different choice of base model. This justifies our choice of a deep learning-based Embedding\&MLP model which is more sophisticated in modeling complex feature interaction. Another benefit of this base model of our choice is that it is general in the sense that many state-of-the-art RS models are developed from the same architectural paradigm. Being able to achieve success on this model implies that similar effects may also be obtained for other deep learning models with similar architectures.
\item Under the proposed ASMG framework, the improvement of ASMG-Linear over IU is minor, and it is sometimes outperformed by other baselines that do not make use of the historical models. This shows that choosing the right design for the meta generator is important, and a simple linear combination is not suitable/sufficient to model the sequential effects. Also, having to drop the models outside of the sliding window limits its ability to accumulate longer-term information.
\item Both ASMG-GRUsingle and ASMG-GRUmulti outperform all the baselines by a significant margin of around $1\%$ in AUC on all four datasets. Note that in the industry, $1\%$ increase in offline AUC is considered very significant \cite{guo2017deepfm,zhou2018deep}, and it is likely to translate to greater improvement in online CTR \cite{cheng2016wide}. The strong results demonstrate the benefits of 1) directly incorporating the historical models and learning by optimizing towards the future (vs SPMF and IncCTR), 2) involving a longer sequence for more informative sequential mining (vs SML), and 3) employing the GRU-based meta generator which is a better design for sequential modeling (vs ASMG-Linear).
\end{itemize}

\begin{table*}[t]
\vspace{-2mm}
\caption{Average AUC \& LogLoss over 7 test periods for four datasets. The results are reported in mean $\pm$ std, computed from 5 runs of model updating experiment. imp\% indicates the relative improvement over the IU baseline.}
\vspace{-2mm}
\centering
\small
\begin{tabular}{ c | c | c | c | c | c | c | c | c}
\hline
\multirow{2}{*}{Method}
& \multicolumn{2}{c|}{Tmall}
& \multicolumn{2}{c|}{Sobazaar}
& \multicolumn{2}{c|}{MovieLens}
& \multicolumn{2}{c}{Lazada} \\
\cline{2-9}
& AUC $\uparrow$ & imp\% & AUC $\uparrow$ & imp\% & AUC $\uparrow$ & imp\% & AUC $\uparrow$ & imp\% \\
\hline
IU
& 0.8180 $\pm$ 0.0007 & -
& 0.7998 $\pm$ 0.0007 & -
& 0.7494 $\pm$ 0.0002 & -
& 0.6381 $\pm$ 0.0001 & - \\
BU-3
& 0.8107 $\pm$ 0.0009 & -0.89\%
& 0.7913 $\pm$ 0.0009 & -1.06\%
& 0.7379 $\pm$ 0.0003 & -1.53\%
& 0.6332 $\pm$ 0.0002 & -0.77\% \\
BU-5
& 0.8002 $\pm$ 0.0009 & -2.18\%
& 0.7824 $\pm$ 0.0012 & -2.18\%
& 0.7280 $\pm$ 0.0005 & -2.86\%
& 0.6287 $\pm$ 0.0004 & -1.47\% \\
BU-7
& 0.7938 $\pm$ 0.0005 & -2.96\%
& 0.7781 $\pm$ 0.0007 & -2.71\%
& 0.7212 $\pm$ 0.0003 & -3.76\%
& 0.6203 $\pm$ 0.0002 & -2.79\% \\
\hline
SPMF
& 0.8187 $\pm$ 0.0006 & 0.09\%
& 0.8007 $\pm$ 0.0004 & 0.11\%
& 0.7511 $\pm$ 0.0002 & 0.23\%
& 0.6381 $\pm$ 0.0002 & 0.00\% \\
IncCTR
& 0.8190 $\pm$ 0.0007 & 0.12\%
& 0.8009 $\pm$ 0.0006 & 0.14\%
& 0.7502 $\pm$ 0.0003 & 0.11\%
& 0.6388 $\pm$ 0.0003 & 0.11\% \\
SML
& 0.8194 $\pm$ 0.0007 & 0.17\%
& 0.8021 $\pm$ 0.0012 & 0.29\%
& 0.7522 $\pm$ 0.0009 & 0.37\%
& 0.6416 $\pm$ 0.0011 & 0.55\% \\
SML-MF
& 0.7628 $\pm$ 0.0013 & -6.75\%
& 0.7782 $\pm$ 0.0017 & -2.70\%
& 0.7242 $\pm$ 0.0012 & -3.36\%
& 0.6100 $\pm$ 0.0016 & -4.40\% \\
\hline
ASMG-Linear
& 0.8190 $\pm$ 0.0006 & 0.12\%
& 0.8002 $\pm$ 0.0008 & 0.05\%
& 0.7524 $\pm$ 0.0002 & 0.40\%
& 0.6390 $\pm$ 0.0001 & 0.14\% \\
ASMG-GRUsingle
& 0.8241 $\pm$ 0.0010 & 0.75\%
& 0.8055 $\pm$ 0.0017 & 0.71\%
& 0.7539 $\pm$ 0.0009 & 0.60\%
& 0.6439 $\pm$ 0.0005 & 0.91\% \\
ASMG-GRUmulti
& \textbf{0.8289 $\pm$ 0.0010} & 1.33\%
& \textbf{0.8108 $\pm$ 0.0017} & 1.38\%
& \textbf{0.7564 $\pm$ 0.0009} & 0.93\%
& \textbf{0.6452 $\pm$ 0.0005} & 1.11\% \\
\hline
& LogLoss $\downarrow$ & imp\% & LogLoss $\downarrow$ & imp\% & LogLoss $\downarrow$ & imp\% & LogLoss $\downarrow$ & imp\% \\
\hline
IU
& 0.5382 $\pm$ 0.0011 & -
& 0.5466 $\pm$ 0.0017 & -
& 0.5871 $\pm$ 0.0002 & -
& 0.4327 $\pm$ 0.0001 & - \\
BU-3
& 0.5518 $\pm$ 0.0009 & -2.53\%
& 0.5536 $\pm$ 0.0013 & -1.28\%
& 0.5949 $\pm$ 0.0004 & -1.33\%
& 0.4342 $\pm$ 0.0001 & -0.35\% \\
BU-5
& 0.5615 $\pm$ 0.0015 & -4.33\%
& 0.5653 $\pm$ 0.0009 & -3.42\%
& 0.6056 $\pm$ 0.0007 & -3.15\%
& 0.4361 $\pm$ 0.0003 & -0.79\% \\
BU-7
& 0.5732 $\pm$ 0.0012 & -6.50\%
& 0.5783 $\pm$ 0.0017 & -5.80\%
& 0.6105 $\pm$ 0.0004 & -3.99\%
& 0.4379 $\pm$ 0.0002 & -1.20\% \\
\hline
SPMF
& 0.5220 $\pm$ 0.0007 & 3.01\%
& 0.5425 $\pm$ 0.0005 & 0.75\%
& 0.5857 $\pm$ 0.0001 & 0.24\%
& 0.4327 $\pm$ 0.0001 & 0.00\% \\
IncCTR
& 0.5344 $\pm$ 0.0010 & 0.71\%
& 0.5458 $\pm$ 0.0007 & 0.15\%
& 0.5865 $\pm$ 0.0003 & 0.10\%
& 0.4321 $\pm$ 0.0001 & 0.14\% \\
SML
& 0.5198 $\pm$ 0.0009 & 3.42\%
& 0.5418 $\pm$ 0.0017 & 0.88\%
& 0.5843 $\pm$ 0.0008 & 0.48\%
& 0.4309 $\pm$ 0.0003 & 0.42\% \\
SML-MF
& 0.5822 $\pm$ 0.0019 & -8.18\%
& 0.5713 $\pm$ 0.0021 & -4.52\%
& 0.6106 $\pm$ 0.0014 & -4.00\%
& 0.4390 $\pm$ 0.0005 & -1.46\% \\
\hline
ASMG-Linear
& 0.5226 $\pm$ 0.0012 & 2.90\%
& 0.5430 $\pm$ 0.0013 & 0.66\%
& 0.5840 $\pm$ 0.0002 & 0.53\%
& 0.4314 $\pm$ 0.0000 & 0.30\% \\
ASMG-GRUsingle
& 0.5154 $\pm$ 0.0018 & 4.24\%
& 0.5370 $\pm$ 0.0017 & 1.76\%
& 0.5828 $\pm$ 0.0011 & 0.73\%
& 0.4304 $\pm$ 0.0003 & 0.53\% \\
ASMG-GRUmulti
& \textbf{0.5079 $\pm$ 0.0018} & 5.63\%
& \textbf{0.5309 $\pm$ 0.0017} & 2.87\%
& \textbf{0.5806 $\pm$ 0.0011} & 1.11\%
& \textbf{0.4299 $\pm$ 0.0003} & 0.65\% \\
\hline
\end{tabular}
\label{tbl:overall_perf}
\vspace{-2mm}
\end{table*}

\begin{figure*}[t]
\centering
\includegraphics[width=1\textwidth]{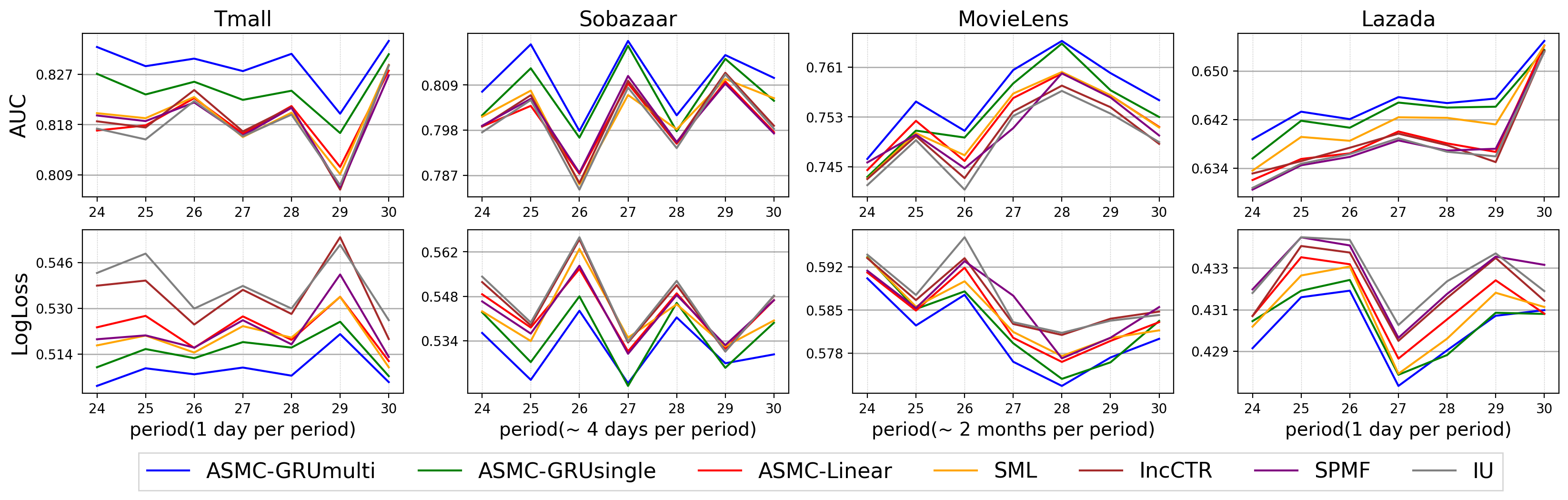}
\vspace{-6mm}
\caption{Prediction performance at each test period for four datasets, averaged over 5 runs of model updating experiment. Note that we omit BU and SML-MF here, as they are not comparable to the rest.}
\label{fig:perf_all}
\vspace{-1mm}
\end{figure*}

Figure \ref{fig:perf_all} shows the disentangled performance at each test period. We can see that both ASMG-GRUsingle and ASMG-GRUmulti consistently outperforms all the other methods at each test period. We purposely divide the periods of different datasets to have different time spans, so that we can test the robustness of our method to different model updating frequencies. Furthermore, we observe that there are drastic fluctuations across periods, which indicates the dynamically changing sequential trends and validates the necessity to adaptively update the meta generator.

\subsection{Ablation Study}
In this section, we focus on ASMG-GRUmulti and test the efficacy of its various designs in regard to the training of GRU meta generator. We compare ASMG-GRUmulti with the following variants, each disables a different design:
\begin{itemize}
\item {\bf ASMG-GRUfull}: This variant trains GRU meta generator on the full sequence of historical models. Illustration of this variant is shown in Appendix \ref{sec:appendix_a} Figure \ref{fig:asmg_grufull}.
\item {\bf ASMG-GRUzero}: This variant uses zero-initialized input hidden state for every period's training instead of continuing on a previously learned hidden state. Illustration of this variant is shown in Appendix \ref{sec:appendix_a} Figure \ref{fig:asmg_gruzero}.
\item {\bf ASMG-GRUunif}: This variant assigns uniform weights to the loss at different time steps. Illustration of this variant is the same as ASMG-GRUmulti, as the weights assigned are not explicitly shown in the diagram.
\item {\bf ASMG-GRUsingle}: This variant performs single-step training at the last output model towards the newly collected data only. Illustration of this variant is shown in Figure \ref{fig:asmg_gru}(b).
\end{itemize}

\subsubsection{Prediction Performance} 
Table \ref{tbl:ablation_perf} presents the prediction performance of four variants. Firstly, by comparing ASMG-GRUmulti with ASMG-GRUfull, we see that they yield comparable performance on MovieLens and Lazada datasets. This demonstrates that ASMG-GRUmulti is able to maintain the performance while achieving better efficiency by training on  truncated sequence. We also observe that for Tmall and Sobazaar, the performance of ASMG-GRUfull slightly drops as compared to ASMG-GRUmulti. This suggests that training on full-length sequence may not always be helpful, as the models at earlier stage may no longer be relevant. Secondly, we see that ASMG-GRUmulti clearly outperforms ASMG-GRUzero. This shows that reusing the previously learned hidden state can help to preserve long-term information. Thirdly, ASMG-GRUmulti also outperforms ASMG-GRUsingle, signifying the benefit of multi-step concurrent training. Finally, ASMG-GRUmulti and ASMG-GRUunif give very similar performance, implying that a simple average of losses may be good enough. The weight decay strategy can be applied on a case-by-case basis.
\subsubsection{Computational Efficiency} 
To show that ASMG-GRUmulti serves to improve efficiency as compared to ASMG-GRUfull, we report the training time of GRU meta generator for both methods at the last test period (i.e., period 30) in Table \ref{tbl:ablation_effcy}. Experiments are conducted using NVIDIA GeForce GTX 1080 Ti with 11GB memory. For ASMG-GRUmulti, we set sequence length $k=3$. For ASMG-GRUfull, the sequence length increases to $20$ at period 30. From the results, we see that the training time of ASMG-GRUfull is about 6.5$\sim$7 times longer than that of ASMG-GRUmulti, consistent with the fact that the training time is linear in sequence length. The results empirically show that fixing the sequence length can greatly improve the computational efficiency, especially at the later training periods.

\begin{table*}[t]
\vspace{-3mm}
\caption{Prediction performance (average AUC \& LogLoss over 7 test periods) of ASMG-GRUmulti and its four variants.}
\vspace{-2mm}
\centering
\small
\begin{tabular}{ c | c | c | c | c | c | c | c | c}
\hline
\multirow{2}{*}{Variant}
& \multicolumn{2}{c|}{Tmall}
& \multicolumn{2}{c|}{Sobazaar}
& \multicolumn{2}{c|}{MovieLens}
& \multicolumn{2}{c}{Lazada} \\
\cline{2-9}
& AUC $\uparrow$ & LogLoss $\downarrow$
& AUC $\uparrow$ & LogLoss $\downarrow$
& AUC $\uparrow$ & LogLoss $\downarrow$
& AUC $\uparrow$ & LogLoss $\downarrow$ \\
\hline
ASMG-GRUfull 
& 0.8267 & 0.5108
& 0.8083 & 0.5323
& \textbf{0.7565} & 0.5811
& 0.6452 & 0.4299 \\
ASMG-GRUzero 
& 0.8224 & 0.5162
& 0.8079 & 0.5343
& 0.7550 & 0.5818
& 0.6440 & 0.4303 \\
ASMG-GRUunif 
& 0.8284 & 0.5102
& 0.8091 & 0.5324
& 0.7563 & 0.5811
& 0.6449 & 0.4300 \\
ASMG-GRUsingle 
& 0.8241 & 0.5154
& 0.8055 & 0.5370
& 0.7539 & 0.5828
& 0.6439 & 0.4304 \\
\hline
ASMG-GRUmulti 
& \textbf{0.8289} & \textbf{0.5079}
& \textbf{0.8108} & \textbf{0.5309}
& 0.7564 & \textbf{0.5806}
& \textbf{0.6452} & \textbf{0.4299} \\
\hline
\end{tabular}
\label{tbl:ablation_perf}
\end{table*}

\begin{table*}[t]
\caption{Training time (in minutes) of GRU meta generator for the two methods at the last test period (i.e., period 30).}
\vspace{-2mm}
\centering
\small
\begin{tabular}{ c | c c c c}
\hline
& Tmall
& Sobazaar
& MovieLens
& Lazada \\
\hline
ASMG-GRUfull & 93.6 & 23.1 & 59.3 & 42.6 \\
ASMG-GRUmulti & 13.8 & 3.4 & 8.7 & 6.2 \\
\hline
\end{tabular}
\label{tbl:ablation_effcy}
\end{table*}

\begin{figure}[t]
\vspace{-2mm}
\subfigure[]{\includegraphics[width=0.49\textwidth]{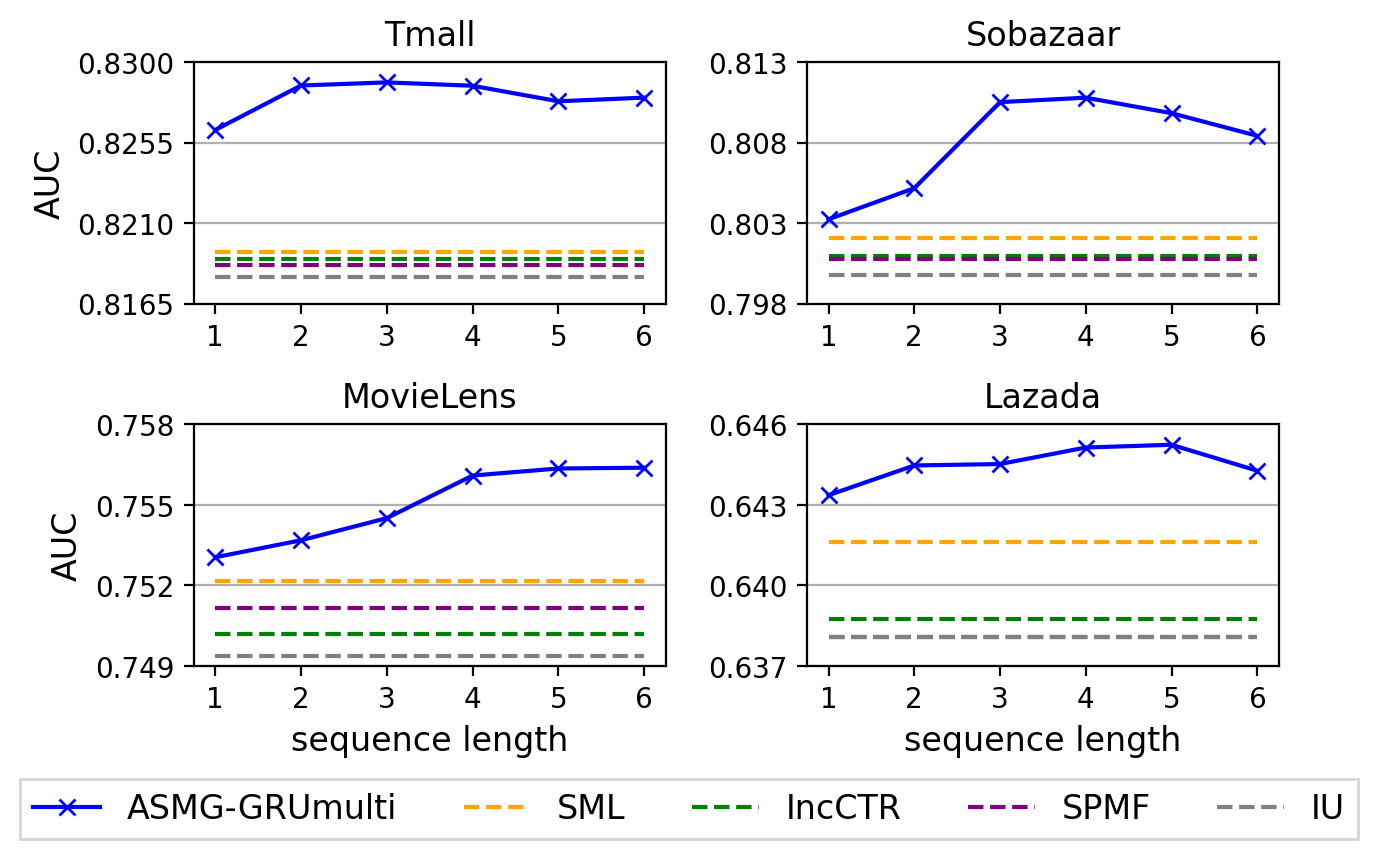}
\label{fig:seq_len_auc}}\vspace{-2mm}
\subfigure[]{\includegraphics[width=0.49\textwidth]{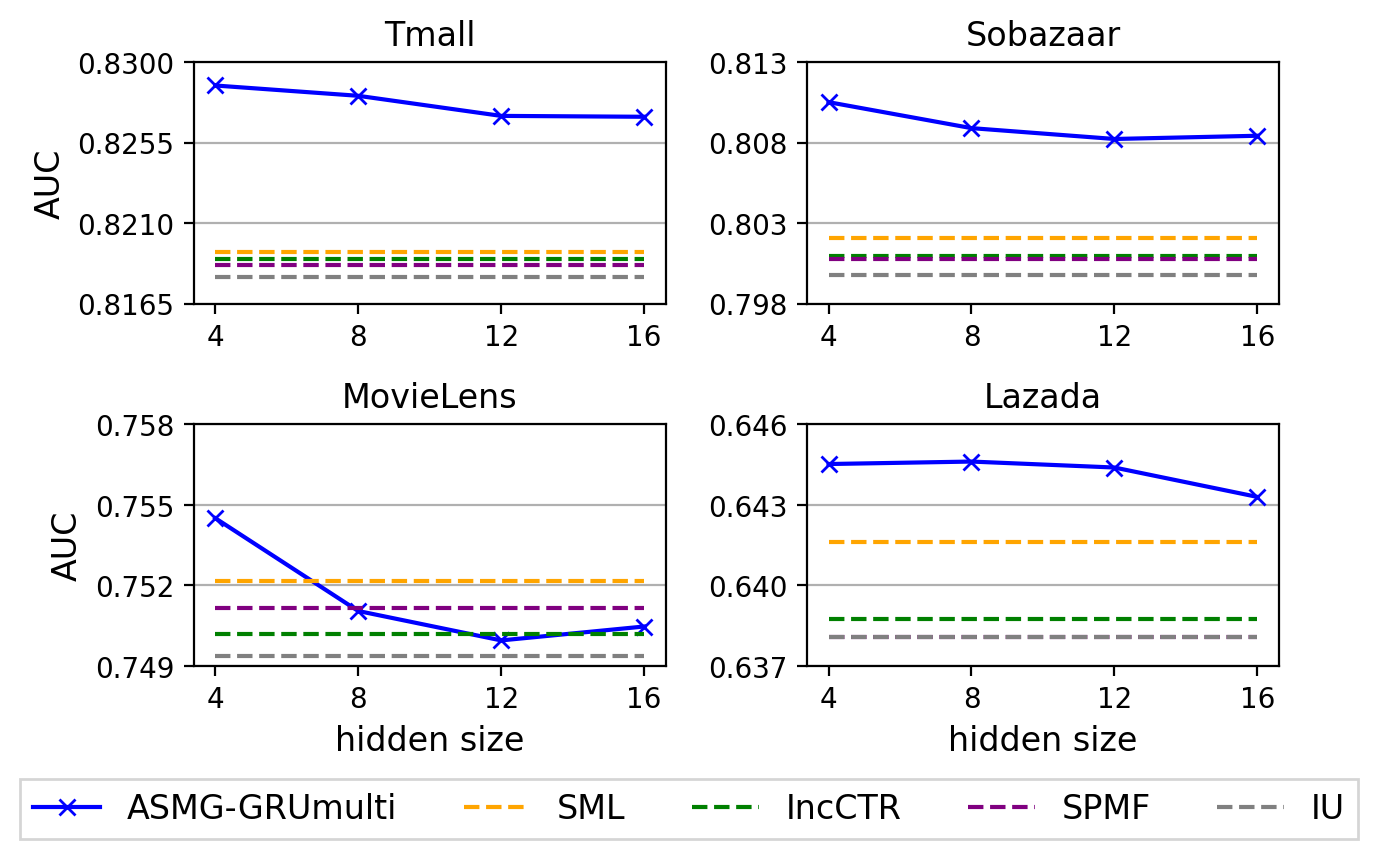}\label{fig:hidden_size_auc}}
\vspace{-4mm}
\caption{Prediction performance (average AUC over 7 test periods) of ASMG-GRUmulti \textit{w.r.t} (a) different input sequence lengths and (b) different GRU hidden sizes.}
\end{figure}

\subsection{Sensitivity Analysis}
We further conduct sensitivity analysis to investigate the effects of two important factors, input sequence length and GRU hidden size, on the performance of ASMG-GRUmulti. Here we only present the analysis in AUC. For results in LogLoss, see Appendix \ref{sec:appendix_b}.

\subsubsection{Effect of Input Sequence Length}
To study the effect of sequence length, we fix the hidden size at 4 and vary the sequence length from 1 to 6. Figure \ref{fig:seq_len_auc} shows the performance of ASMG-GRUmulti \textit{w.r.t} different sequence lengths. Firstly, we see that ASMG-GRUmulti is able to outperform all the baselines regardless of sequence length. Even for $k=1$ (i.e., applying ASMG-GRUmulti on the most recent model only), ASMG-GRUmulti is still slightly better than the strongest baseline SML, which is a model-based method that considers transfer between 2 periods (i.e., sequence length is 2). This can be attributed to the GRU design which better models the sequential patterns, and also the strategy to continue from a previously learned hidden state, which helps retain longer memory. Secondly, we observe from the trends that as long as the sequence length is long enough for effective training of GRUs (e.g., the minimum length should be 2 for Tmall and Lazada, 3 for Sobazaar, and 4 for MovieLens), the performance tends to stabilize at a satisfactory level. However, it is also noticed that further increasing the sequence length may degrade the performance (e.g., downward trend observed for Sobazaar after 4 and Lazada after 5). Longer sequence length is also not desirable due to the expensive computation. Therefore, the optimal range of sequence length should be from 3 to 5.

\subsubsection{Effect of GRU Hidden Size}
We also study the effect of GRU hidden size on the performance of ASMG-GRUmulti. By fixing the sequence length at 3, we vary the hidden size in $\{4, 8, 12, 16\}$. The results are shown in Figure \ref{fig:hidden_size_auc}. We can see that the performance generally drops for more than 4 hidden units. For MovieLens, the degradation is so severe that the performance becomes worse than some of the baselines. This implies that overfitting may occur if the meta generator is over-parameterized by increasing the GRU hidden size. Furthermore, complex GRU meta generator is also expensive to train. Hence, we suggest that setting the hidden size to 4 is a good starting point.

\section{Conclusion}
In this work, we study the model updating strategies for RSs, aiming to achieve recency with incremental update while still being able to prevent forgetting of past knowledge. To this end, we propose a novel ASMG framework, which produces a better serving model by encoding a sequence of incrementally updated models in the past via a meta generator. We introduce a meta generator design based on GRUs and further develop some strategies to improve its training efficiency and sequential modeling ability. By instantiating the proposed method on a general deep learning-based RS model, we demonstrate that it achieves the state of the art on four real-world datasets. In the future, we will investigate more effective meta generator designs considering different base model architectures.

\begin{acks}
This research is supported, in part, by Alibaba Group through Alibaba Innovative Research (AIR) Program and Alibaba-NTU Singapore Joint Research Institute (JRI), Nanyang Technological University, Singapore.
\end{acks}

\newpage
\bibliographystyle{ACM-Reference-Format}
\bibliography{reference}


\begin{thebibliography}{31}


\ifx \showCODEN    \undefined \def \showCODEN     #1{\unskip}     \fi
\ifx \showDOI      \undefined \def \showDOI       #1{#1}\fi
\ifx \showISBNx    \undefined \def \showISBNx     #1{\unskip}     \fi
\ifx \showISBNxiii \undefined \def \showISBNxiii  #1{\unskip}     \fi
\ifx \showISSN     \undefined \def \showISSN      #1{\unskip}     \fi
\ifx \showLCCN     \undefined \def \showLCCN      #1{\unskip}     \fi
\ifx \shownote     \undefined \def \shownote      #1{#1}          \fi
\ifx \showarticletitle \undefined \def \showarticletitle #1{#1}   \fi
\ifx \showURL      \undefined \def \showURL       {\relax}        \fi
\providecommand\bibfield[2]{#2}
\providecommand\bibinfo[2]{#2}
\providecommand\natexlab[1]{#1}
\providecommand\showeprint[2][]{arXiv:#2}

\bibitem[\protect\citeauthoryear{Chen, Yin, Yao, and Cui}{Chen
  et~al\mbox{.}}{2013}]%
        {chen2013terec}
\bibfield{author}{\bibinfo{person}{Chen Chen}, \bibinfo{person}{Hongzhi Yin},
  \bibinfo{person}{Junjie Yao}, {and} \bibinfo{person}{Bin Cui}.}
  \bibinfo{year}{2013}\natexlab{}.
\newblock \showarticletitle{Terec: A temporal recommender system over tweet
  stream}.
\newblock \bibinfo{journal}{\emph{Proceedings of the VLDB Endowment}}
  \bibinfo{volume}{6}, \bibinfo{number}{12} (\bibinfo{year}{2013}),
  \bibinfo{pages}{1254--1257}.
\newblock


\bibitem[\protect\citeauthoryear{Cheng, Koc, Harmsen, Shaked, Chandra, Aradhye,
  Anderson, Corrado, Chai, Ispir, et~al\mbox{.}}{Cheng et~al\mbox{.}}{2016}]%
        {cheng2016wide}
\bibfield{author}{\bibinfo{person}{Heng-Tze Cheng}, \bibinfo{person}{Levent
  Koc}, \bibinfo{person}{Jeremiah Harmsen}, \bibinfo{person}{Tal Shaked},
  \bibinfo{person}{Tushar Chandra}, \bibinfo{person}{Hrishi Aradhye},
  \bibinfo{person}{Glen Anderson}, \bibinfo{person}{Greg Corrado},
  \bibinfo{person}{Wei Chai}, \bibinfo{person}{Mustafa Ispir}, {et~al\mbox{.}}}
  \bibinfo{year}{2016}\natexlab{}.
\newblock \showarticletitle{Wide \& Deep Learning for Recommender Systems}. In
  \bibinfo{booktitle}{\emph{DLRS@ RecSys}}.
\newblock


\bibitem[\protect\citeauthoryear{Chung, Gulcehre, Cho, and Bengio}{Chung
  et~al\mbox{.}}{2014}]%
        {chung2014empirical}
\bibfield{author}{\bibinfo{person}{Junyoung Chung}, \bibinfo{person}{Caglar
  Gulcehre}, \bibinfo{person}{Kyunghyun Cho}, {and} \bibinfo{person}{Yoshua
  Bengio}.} \bibinfo{year}{2014}\natexlab{}.
\newblock \showarticletitle{Empirical evaluation of gated recurrent neural
  networks on sequence modeling}. In \bibinfo{booktitle}{\emph{NIPS 2014
  Workshop on Deep Learning, December 2014}}.
\newblock


\bibitem[\protect\citeauthoryear{Devooght, Kourtellis, and Mantrach}{Devooght
  et~al\mbox{.}}{2015}]%
        {devooght2015dynamic}
\bibfield{author}{\bibinfo{person}{Robin Devooght}, \bibinfo{person}{Nicolas
  Kourtellis}, {and} \bibinfo{person}{Amin Mantrach}.}
  \bibinfo{year}{2015}\natexlab{}.
\newblock \showarticletitle{Dynamic matrix factorization with priors on unknown
  values}. In \bibinfo{booktitle}{\emph{Proceedings of the 21th ACM SIGKDD
  international conference on knowledge discovery and data mining}}.
  \bibinfo{pages}{189--198}.
\newblock


\bibitem[\protect\citeauthoryear{Diaz-Aviles, Drumond, Schmidt-Thieme, and
  Nejdl}{Diaz-Aviles et~al\mbox{.}}{2012}]%
        {diaz2012real}
\bibfield{author}{\bibinfo{person}{Ernesto Diaz-Aviles}, \bibinfo{person}{Lucas
  Drumond}, \bibinfo{person}{Lars Schmidt-Thieme}, {and}
  \bibinfo{person}{Wolfgang Nejdl}.} \bibinfo{year}{2012}\natexlab{}.
\newblock \showarticletitle{Real-time top-n recommendation in social streams}.
  In \bibinfo{booktitle}{\emph{Proceedings of the sixth ACM conference on
  Recommender systems}}. \bibinfo{pages}{59--66}.
\newblock


\bibitem[\protect\citeauthoryear{Frig{\'o}, P{\'a}lovics, Kelen, Kocsis, and
  Bencz{\'u}r}{Frig{\'o} et~al\mbox{.}}{2017}]%
        {frigo2017online}
\bibfield{author}{\bibinfo{person}{Erzs{\'e}bet Frig{\'o}},
  \bibinfo{person}{R{\'o}bert P{\'a}lovics}, \bibinfo{person}{Domokos Kelen},
  \bibinfo{person}{Levente Kocsis}, {and} \bibinfo{person}{Andr{\'a}s
  Bencz{\'u}r}.} \bibinfo{year}{2017}\natexlab{}.
\newblock \showarticletitle{Online ranking prediction in non-stationary
  environments}.
\newblock  (\bibinfo{year}{2017}).
\newblock


\bibitem[\protect\citeauthoryear{Guo, Tang, Ye, Li, and He}{Guo
  et~al\mbox{.}}{2017}]%
        {guo2017deepfm}
\bibfield{author}{\bibinfo{person}{Huifeng Guo}, \bibinfo{person}{Ruiming
  Tang}, \bibinfo{person}{Yunming Ye}, \bibinfo{person}{Zhenguo Li}, {and}
  \bibinfo{person}{Xiuqiang He}.} \bibinfo{year}{2017}\natexlab{}.
\newblock \showarticletitle{DeepFM: a factorization-machine based neural
  network for CTR prediction}. In \bibinfo{booktitle}{\emph{Proceedings of the
  26th International Joint Conference on Artificial Intelligence}}.
  \bibinfo{pages}{1725--1731}.
\newblock


\bibitem[\protect\citeauthoryear{Guo, Yin, Wang, Chen, Zhou, and Quoc
  Viet~Hung}{Guo et~al\mbox{.}}{2019}]%
        {guo2019streaming}
\bibfield{author}{\bibinfo{person}{Lei Guo}, \bibinfo{person}{Hongzhi Yin},
  \bibinfo{person}{Qinyong Wang}, \bibinfo{person}{Tong Chen},
  \bibinfo{person}{Alexander Zhou}, {and} \bibinfo{person}{Nguyen Quoc
  Viet~Hung}.} \bibinfo{year}{2019}\natexlab{}.
\newblock \showarticletitle{Streaming session-based recommendation}. In
  \bibinfo{booktitle}{\emph{Proceedings of the 25th ACM SIGKDD International
  Conference on Knowledge Discovery \& Data Mining}}.
  \bibinfo{pages}{1569--1577}.
\newblock


\bibitem[\protect\citeauthoryear{Jugovac, Jannach, and Karimi}{Jugovac
  et~al\mbox{.}}{2018}]%
        {jugovac2018streamingrec}
\bibfield{author}{\bibinfo{person}{Michael Jugovac}, \bibinfo{person}{Dietmar
  Jannach}, {and} \bibinfo{person}{Mozhgan Karimi}.}
  \bibinfo{year}{2018}\natexlab{}.
\newblock \showarticletitle{Streamingrec: a framework for benchmarking
  stream-based news recommenders}. In \bibinfo{booktitle}{\emph{Proceedings of
  the 12th ACM Conference on Recommender Systems}}. \bibinfo{pages}{269--273}.
\newblock


\bibitem[\protect\citeauthoryear{Kirkpatrick, Pascanu, Rabinowitz, Veness,
  Desjardins, Rusu, Milan, Quan, Ramalho, Grabska-Barwinska,
  et~al\mbox{.}}{Kirkpatrick et~al\mbox{.}}{2017}]%
        {kirkpatrick2017overcoming}
\bibfield{author}{\bibinfo{person}{James Kirkpatrick}, \bibinfo{person}{Razvan
  Pascanu}, \bibinfo{person}{Neil Rabinowitz}, \bibinfo{person}{Joel Veness},
  \bibinfo{person}{Guillaume Desjardins}, \bibinfo{person}{Andrei~A Rusu},
  \bibinfo{person}{Kieran Milan}, \bibinfo{person}{John Quan},
  \bibinfo{person}{Tiago Ramalho}, \bibinfo{person}{Agnieszka
  Grabska-Barwinska}, {et~al\mbox{.}}} \bibinfo{year}{2017}\natexlab{}.
\newblock \showarticletitle{Overcoming catastrophic forgetting in neural
  networks}.
\newblock \bibinfo{journal}{\emph{Proceedings of the national academy of
  sciences}} \bibinfo{volume}{114}, \bibinfo{number}{13}
  (\bibinfo{year}{2017}), \bibinfo{pages}{3521--3526}.
\newblock


\bibitem[\protect\citeauthoryear{Mallya and Lazebnik}{Mallya and
  Lazebnik}{2018}]%
        {mallya2018packnet}
\bibfield{author}{\bibinfo{person}{Arun Mallya} {and} \bibinfo{person}{Svetlana
  Lazebnik}.} \bibinfo{year}{2018}\natexlab{}.
\newblock \showarticletitle{Packnet: Adding multiple tasks to a single network
  by iterative pruning}. In \bibinfo{booktitle}{\emph{Proceedings of the IEEE
  Conference on Computer Vision and Pattern Recognition}}.
  \bibinfo{pages}{7765--7773}.
\newblock


\bibitem[\protect\citeauthoryear{Mi and Faltings}{Mi and Faltings}{2020}]%
        {mi2020memory}
\bibfield{author}{\bibinfo{person}{Fei Mi} {and} \bibinfo{person}{Boi
  Faltings}.} \bibinfo{year}{2020}\natexlab{}.
\newblock \showarticletitle{Memory Augmented Neural Model for Incremental
  Session-based Recommendation}.
\newblock \bibinfo{journal}{\emph{arXiv preprint arXiv:2005.01573}}
  (\bibinfo{year}{2020}).
\newblock


\bibitem[\protect\citeauthoryear{Mi, Lin, and Faltings}{Mi
  et~al\mbox{.}}{2020}]%
        {mi2020ader}
\bibfield{author}{\bibinfo{person}{Fei Mi}, \bibinfo{person}{Xiaoyu Lin}, {and}
  \bibinfo{person}{Boi Faltings}.} \bibinfo{year}{2020}\natexlab{}.
\newblock \showarticletitle{Ader: Adaptively distilled exemplar replay towards
  continual learning for session-based recommendation}. In
  \bibinfo{booktitle}{\emph{Fourteenth ACM Conference on Recommender Systems}}.
  \bibinfo{pages}{408--413}.
\newblock


\bibitem[\protect\citeauthoryear{Papagelis, Rousidis, Plexousakis, and
  Theoharopoulos}{Papagelis et~al\mbox{.}}{2005}]%
        {papagelis2005incremental}
\bibfield{author}{\bibinfo{person}{Manos Papagelis}, \bibinfo{person}{Ioannis
  Rousidis}, \bibinfo{person}{Dimitris Plexousakis}, {and}
  \bibinfo{person}{Elias Theoharopoulos}.} \bibinfo{year}{2005}\natexlab{}.
\newblock \showarticletitle{Incremental collaborative filtering for
  highly-scalable recommendation algorithms}. In
  \bibinfo{booktitle}{\emph{International Symposium on Methodologies for
  Intelligent Systems}}. Springer, \bibinfo{pages}{553--561}.
\newblock


\bibitem[\protect\citeauthoryear{Qiu, Yin, Huang, and Chen}{Qiu
  et~al\mbox{.}}{2020}]%
        {qiu2020gag}
\bibfield{author}{\bibinfo{person}{Ruihong Qiu}, \bibinfo{person}{Hongzhi Yin},
  \bibinfo{person}{Zi Huang}, {and} \bibinfo{person}{Tong Chen}.}
  \bibinfo{year}{2020}\natexlab{}.
\newblock \showarticletitle{Gag: Global attributed graph neural network for
  streaming session-based recommendation}. In
  \bibinfo{booktitle}{\emph{Proceedings of the 43rd International ACM SIGIR
  Conference on Research and Development in Information Retrieval}}.
  \bibinfo{pages}{669--678}.
\newblock


\bibitem[\protect\citeauthoryear{Rebuffi, Kolesnikov, Sperl, and
  Lampert}{Rebuffi et~al\mbox{.}}{2017}]%
        {rebuffi2017icarl}
\bibfield{author}{\bibinfo{person}{Sylvestre-Alvise Rebuffi},
  \bibinfo{person}{Alexander Kolesnikov}, \bibinfo{person}{Georg Sperl}, {and}
  \bibinfo{person}{Christoph~H Lampert}.} \bibinfo{year}{2017}\natexlab{}.
\newblock \showarticletitle{icarl: Incremental classifier and representation
  learning}. In \bibinfo{booktitle}{\emph{Proceedings of the IEEE conference on
  Computer Vision and Pattern Recognition}}. \bibinfo{pages}{2001--2010}.
\newblock


\bibitem[\protect\citeauthoryear{Rendle and Schmidt-Thieme}{Rendle and
  Schmidt-Thieme}{2008}]%
        {rendle2008online}
\bibfield{author}{\bibinfo{person}{Steffen Rendle} {and} \bibinfo{person}{Lars
  Schmidt-Thieme}.} \bibinfo{year}{2008}\natexlab{}.
\newblock \showarticletitle{Online-updating regularized kernel matrix
  factorization models for large-scale recommender systems}. In
  \bibinfo{booktitle}{\emph{Proceedings of the 2008 ACM conference on
  Recommender systems}}. \bibinfo{pages}{251--258}.
\newblock


\bibitem[\protect\citeauthoryear{Robins}{Robins}{1995}]%
        {robins1995catastrophic}
\bibfield{author}{\bibinfo{person}{Anthony Robins}.}
  \bibinfo{year}{1995}\natexlab{}.
\newblock \showarticletitle{Catastrophic forgetting, rehearsal and
  pseudorehearsal}.
\newblock \bibinfo{journal}{\emph{Connection Science}} \bibinfo{volume}{7},
  \bibinfo{number}{2} (\bibinfo{year}{1995}), \bibinfo{pages}{123--146}.
\newblock


\bibitem[\protect\citeauthoryear{Rusu, Rabinowitz, Desjardins, Soyer,
  Kirkpatrick, Kavukcuoglu, Pascanu, and Hadsell}{Rusu et~al\mbox{.}}{2016}]%
        {rusu2016progressive}
\bibfield{author}{\bibinfo{person}{Andrei~A Rusu}, \bibinfo{person}{Neil~C
  Rabinowitz}, \bibinfo{person}{Guillaume Desjardins}, \bibinfo{person}{Hubert
  Soyer}, \bibinfo{person}{James Kirkpatrick}, \bibinfo{person}{Koray
  Kavukcuoglu}, \bibinfo{person}{Razvan Pascanu}, {and} \bibinfo{person}{Raia
  Hadsell}.} \bibinfo{year}{2016}\natexlab{}.
\newblock \showarticletitle{Progressive neural networks}.
\newblock \bibinfo{journal}{\emph{arXiv preprint arXiv:1606.04671}}
  (\bibinfo{year}{2016}).
\newblock


\bibitem[\protect\citeauthoryear{Shan, Hoens, Jiao, Wang, Yu, and Mao}{Shan
  et~al\mbox{.}}{2016}]%
        {shan2016deep}
\bibfield{author}{\bibinfo{person}{Ying Shan}, \bibinfo{person}{T~Ryan Hoens},
  \bibinfo{person}{Jian Jiao}, \bibinfo{person}{Haijing Wang},
  \bibinfo{person}{Dong Yu}, {and} \bibinfo{person}{JC Mao}.}
  \bibinfo{year}{2016}\natexlab{}.
\newblock \showarticletitle{Deep crossing: Web-scale modeling without manually
  crafted combinatorial features}. In \bibinfo{booktitle}{\emph{Proceedings of
  the 22nd ACM SIGKDD international conference on knowledge discovery and data
  mining}}. \bibinfo{pages}{255--262}.
\newblock


\bibitem[\protect\citeauthoryear{Shin, Lee, Kim, and Kim}{Shin
  et~al\mbox{.}}{2017}]%
        {shin2017continual}
\bibfield{author}{\bibinfo{person}{Hanul Shin}, \bibinfo{person}{Jung~Kwon
  Lee}, \bibinfo{person}{Jaehong Kim}, {and} \bibinfo{person}{Jiwon Kim}.}
  \bibinfo{year}{2017}\natexlab{}.
\newblock \showarticletitle{Continual learning with deep generative replay}. In
  \bibinfo{booktitle}{\emph{Advances in neural information processing
  systems}}. \bibinfo{pages}{2990--2999}.
\newblock


\bibitem[\protect\citeauthoryear{Vinagre, Jorge, and Gama}{Vinagre
  et~al\mbox{.}}{2014}]%
        {vinagre2014fast}
\bibfield{author}{\bibinfo{person}{Jo{\~a}o Vinagre},
  \bibinfo{person}{Al{\'\i}pio~M{\'a}rio Jorge}, {and}
  \bibinfo{person}{Jo{\~a}o Gama}.} \bibinfo{year}{2014}\natexlab{}.
\newblock \showarticletitle{Fast incremental matrix factorization for
  recommendation with positive-only feedback}. In
  \bibinfo{booktitle}{\emph{International Conference on User Modeling,
  Adaptation, and Personalization}}. Springer, \bibinfo{pages}{459--470}.
\newblock


\bibitem[\protect\citeauthoryear{Wang, Huang, Wu, Guo, and Lan}{Wang
  et~al\mbox{.}}{2016}]%
        {wang2016incremental}
\bibfield{author}{\bibinfo{person}{JianGuo Wang},
  \bibinfo{person}{Joshua~Zhexue Huang}, \bibinfo{person}{Dingming Wu},
  \bibinfo{person}{Jiafeng Guo}, {and} \bibinfo{person}{Yanyan Lan}.}
  \bibinfo{year}{2016}\natexlab{}.
\newblock \showarticletitle{An incremental model on search engine query
  recommendation}.
\newblock \bibinfo{journal}{\emph{Neurocomputing}}  \bibinfo{volume}{218}
  (\bibinfo{year}{2016}), \bibinfo{pages}{423--431}.
\newblock


\bibitem[\protect\citeauthoryear{Wang, Yin, Huang, Wang, Du, and Nguyen}{Wang
  et~al\mbox{.}}{2018}]%
        {wang2018streaming}
\bibfield{author}{\bibinfo{person}{Weiqing Wang}, \bibinfo{person}{Hongzhi
  Yin}, \bibinfo{person}{Zi Huang}, \bibinfo{person}{Qinyong Wang},
  \bibinfo{person}{Xingzhong Du}, {and} \bibinfo{person}{Quoc Viet~Hung
  Nguyen}.} \bibinfo{year}{2018}\natexlab{}.
\newblock \showarticletitle{Streaming ranking based recommender systems}. In
  \bibinfo{booktitle}{\emph{The 41st International ACM SIGIR Conference on
  Research \& Development in Information Retrieval}}.
  \bibinfo{pages}{525--534}.
\newblock


\bibitem[\protect\citeauthoryear{Wang, Guo, Tang, Liu, and He}{Wang
  et~al\mbox{.}}{2020}]%
        {wang2020practical}
\bibfield{author}{\bibinfo{person}{Yichao Wang}, \bibinfo{person}{Huifeng Guo},
  \bibinfo{person}{Ruiming Tang}, \bibinfo{person}{Zhirong Liu}, {and}
  \bibinfo{person}{Xiuqiang He}.} \bibinfo{year}{2020}\natexlab{}.
\newblock \showarticletitle{A Practical Incremental Method to Train Deep CTR
  Models}.
\newblock \bibinfo{journal}{\emph{arXiv preprint arXiv:2009.02147}}
  (\bibinfo{year}{2020}).
\newblock


\bibitem[\protect\citeauthoryear{Wen, Cao, and Huang}{Wen
  et~al\mbox{.}}{2018}]%
        {wen2018few}
\bibfield{author}{\bibinfo{person}{Junfeng Wen}, \bibinfo{person}{Yanshuai
  Cao}, {and} \bibinfo{person}{Ruitong Huang}.}
  \bibinfo{year}{2018}\natexlab{}.
\newblock \showarticletitle{Few-shot self reminder to overcome catastrophic
  forgetting}.
\newblock \bibinfo{journal}{\emph{arXiv preprint arXiv:1812.00543}}
  (\bibinfo{year}{2018}).
\newblock


\bibitem[\protect\citeauthoryear{Xu, Zhang, Guo, Guo, Tang, and Coates}{Xu
  et~al\mbox{.}}{2020}]%
        {xu2020graphsail}
\bibfield{author}{\bibinfo{person}{Yishi Xu}, \bibinfo{person}{Yingxue Zhang},
  \bibinfo{person}{Wei Guo}, \bibinfo{person}{Huifeng Guo},
  \bibinfo{person}{Ruiming Tang}, {and} \bibinfo{person}{Mark Coates}.}
  \bibinfo{year}{2020}\natexlab{}.
\newblock \showarticletitle{GraphSAIL: Graph Structure Aware Incremental
  Learning for Recommender Systems}. In \bibinfo{booktitle}{\emph{Proceedings
  of the 29th ACM International Conference on Information \& Knowledge
  Management}}. \bibinfo{pages}{2861--2868}.
\newblock


\bibitem[\protect\citeauthoryear{Zenke, Poole, and Ganguli}{Zenke
  et~al\mbox{.}}{2017}]%
        {zenke2017continual}
\bibfield{author}{\bibinfo{person}{Friedemann Zenke}, \bibinfo{person}{Ben
  Poole}, {and} \bibinfo{person}{Surya Ganguli}.}
  \bibinfo{year}{2017}\natexlab{}.
\newblock \showarticletitle{Continual learning through synaptic intelligence}.
  In \bibinfo{booktitle}{\emph{International Conference on Machine Learning}}.
  PMLR, \bibinfo{pages}{3987--3995}.
\newblock


\bibitem[\protect\citeauthoryear{Zhang, Feng, Wang, He, Wang, Li, and
  Zhang}{Zhang et~al\mbox{.}}{2020}]%
        {zhang2020retrain}
\bibfield{author}{\bibinfo{person}{Yang Zhang}, \bibinfo{person}{Fuli Feng},
  \bibinfo{person}{Chenxu Wang}, \bibinfo{person}{Xiangnan He},
  \bibinfo{person}{Meng Wang}, \bibinfo{person}{Yan Li}, {and}
  \bibinfo{person}{Yongdong Zhang}.} \bibinfo{year}{2020}\natexlab{}.
\newblock \showarticletitle{How to retrain recommender system? A sequential
  meta-learning method}. In \bibinfo{booktitle}{\emph{Proceedings of the 43rd
  International ACM SIGIR Conference on Research and Development in Information
  Retrieval}}. \bibinfo{pages}{1479--1488}.
\newblock


\bibitem[\protect\citeauthoryear{Zhao, Wang, Wang, and Liu}{Zhao
  et~al\mbox{.}}{2020}]%
        {zhao2020stratified}
\bibfield{author}{\bibinfo{person}{Yan Zhao}, \bibinfo{person}{Shoujin Wang},
  \bibinfo{person}{Yan Wang}, {and} \bibinfo{person}{Hongwei Liu}.}
  \bibinfo{year}{2020}\natexlab{}.
\newblock \showarticletitle{Stratified and time-aware sampling based adaptive
  ensemble learning for streaming recommendations}.
\newblock \bibinfo{journal}{\emph{Applied Intelligence}}
  (\bibinfo{year}{2020}), \bibinfo{pages}{1--21}.
\newblock


\bibitem[\protect\citeauthoryear{Zhou, Zhu, Song, Fan, Zhu, Ma, Yan, Jin, Li,
  and Gai}{Zhou et~al\mbox{.}}{2018}]%
        {zhou2018deep}
\bibfield{author}{\bibinfo{person}{Guorui Zhou}, \bibinfo{person}{Xiaoqiang
  Zhu}, \bibinfo{person}{Chenru Song}, \bibinfo{person}{Ying Fan},
  \bibinfo{person}{Han Zhu}, \bibinfo{person}{Xiao Ma},
  \bibinfo{person}{Yanghui Yan}, \bibinfo{person}{Junqi Jin},
  \bibinfo{person}{Han Li}, {and} \bibinfo{person}{Kun Gai}.}
  \bibinfo{year}{2018}\natexlab{}.
\newblock \showarticletitle{Deep interest network for click-through rate
  prediction}. In \bibinfo{booktitle}{\emph{Proceedings of the 24th ACM SIGKDD
  International Conference on Knowledge Discovery \& Data Mining}}.
  \bibinfo{pages}{1059--1068}.
\newblock


\end{thebibliography}


\appendix
\newpage
\section{Illustrations of Variants}
\label{sec:appendix_a}
\begin{figure}[h]
\subfigure[]{\includegraphics[width=0.49\textwidth]{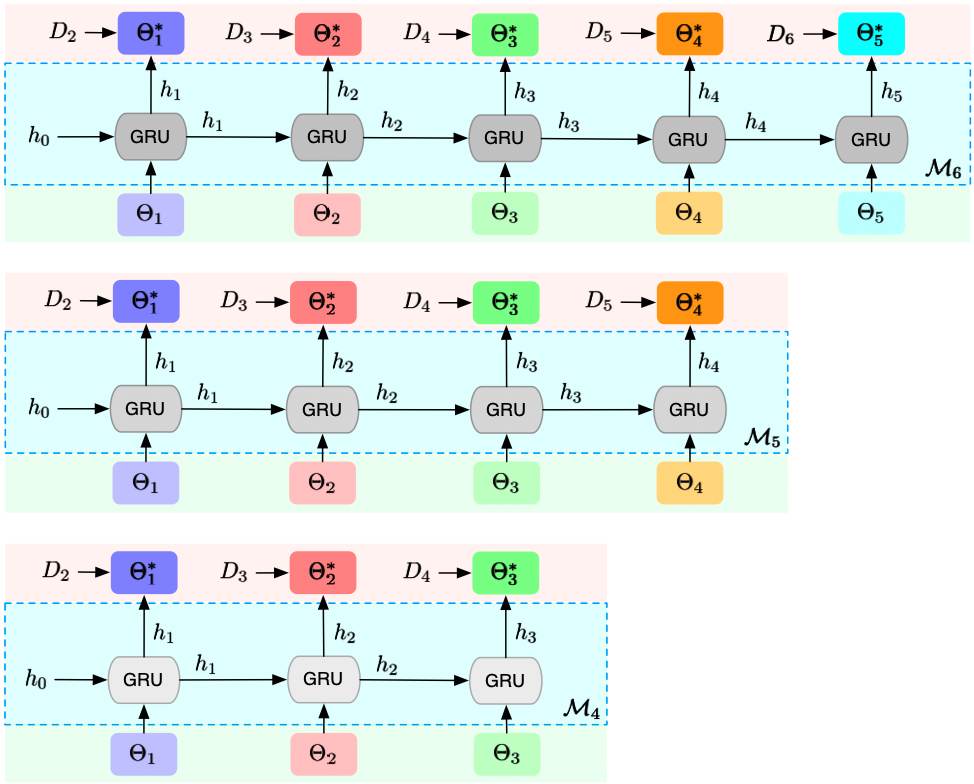}\label{fig:asmg_grufull}}
\subfigure[]{\includegraphics[width=0.49\textwidth]{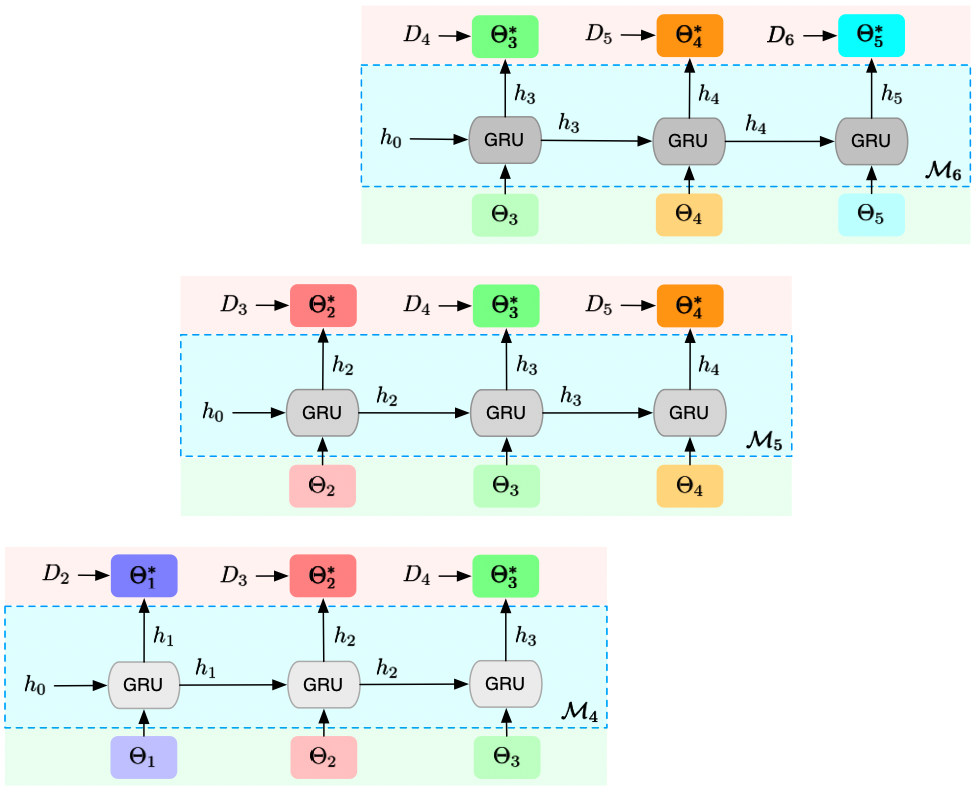}\label{fig:asmg_gruzero}}
\caption{(a) \textbf{ASMG-GRUfull} trains GRU meta generator on the full sequence of historical models. (b) \textbf{ASMG-GRUzero} uses zero-initialized input hidden state for every period's training (i.e., same input hidden state $h_0$ for the training of $\mathcal{M}_4$, $\mathcal{M}_5$ and $\mathcal{M}_6$) instead of continuing on a previously learned hidden state.}
\end{figure}
\vspace{18mm}

\section{Sensitivity Analysis in LogLoss}
\label{sec:appendix_b}
\begin{figure}[h]
\subfigure[]{\includegraphics[width=0.49\textwidth]{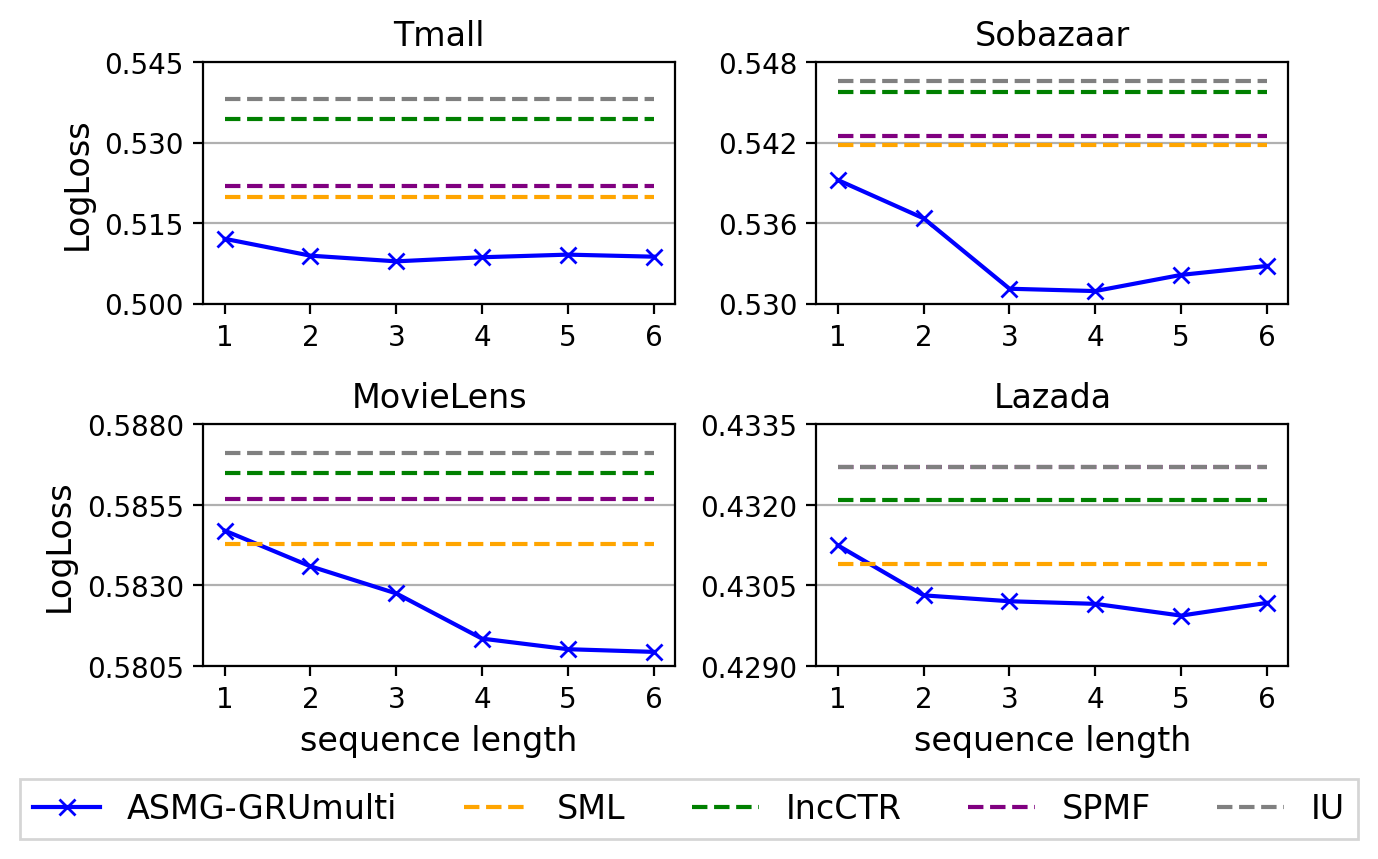}}
\subfigure[]{\includegraphics[width=0.49\textwidth]{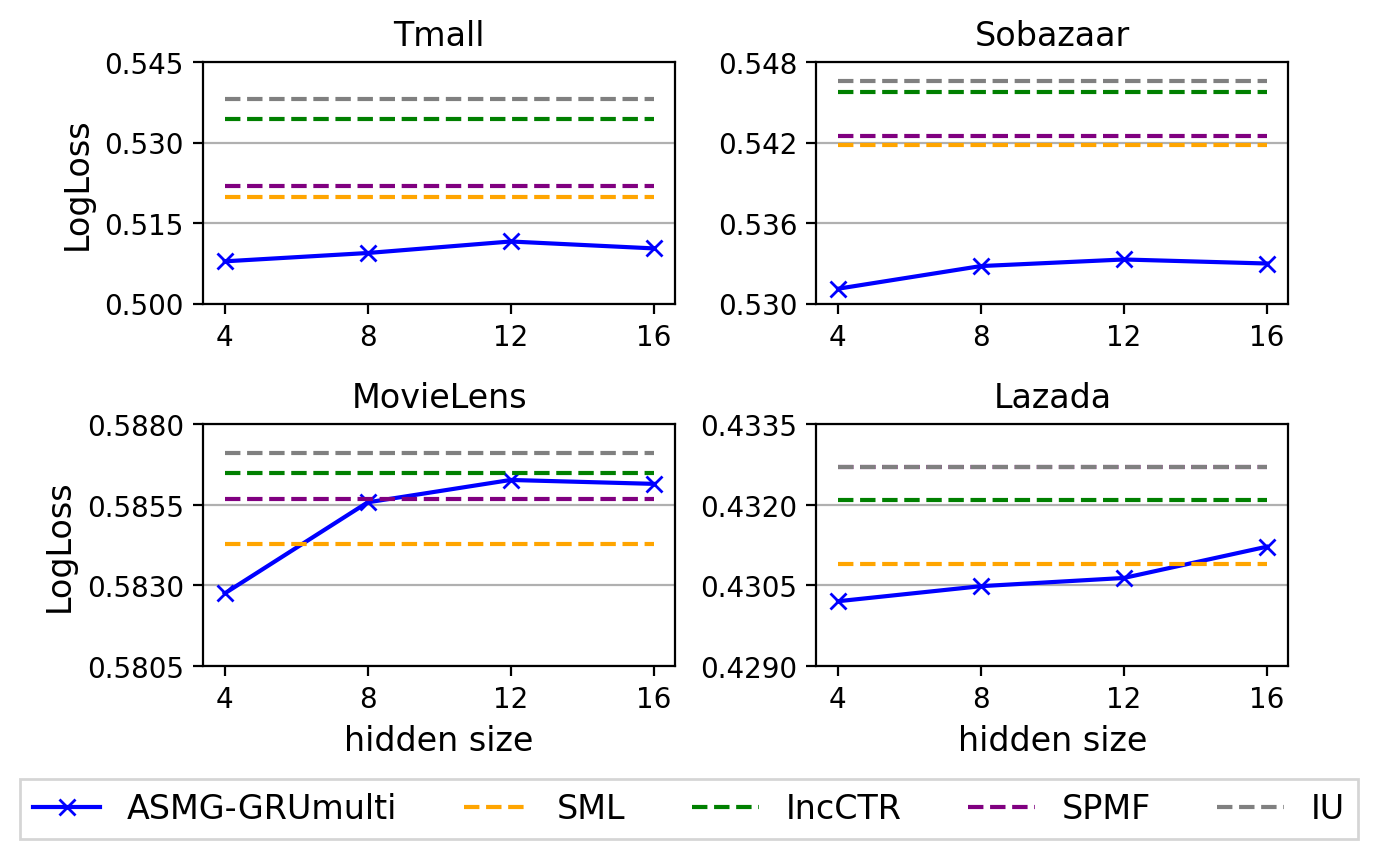}}
\caption{Prediction performance (average LogLoss over 7 test periods) of ASMG-GRUmulti \textit{w.r.t} (a) different input sequence lengths and (b) different GRU hidden sizes.}
\label{fig:sens_logloss}
\end{figure}

\end{document}